\documentclass[12pt]{article}  
\usepackage{epsfig}  
  
\newcommand{\be}{\begin{equation}}  
\newcommand{\ee}{\end{equation}}  
\newcommand{\bea}{\begin{eqnarray}}  
\newcommand{\eea}{\end{eqnarray}}

\newcommand{\parp}{\partial}  
\newcommand{\vet}{\overline}  
  
\begin{document}  
\bibliographystyle{unsrt}

\title{  
Universality Classes of Self-Avoiding Fixed-Connectivity Membranes}  
\author{\small \\ Mark~J. Bowick$^{(1)}$\hspace{-.16cm} 
\footnote{\tt bowick@physics.syr.edu \hspace{.3cm} 
 {$^{\ddagger}$}cacciuto@physics.syr.edu}\ , 
Angelo Cacciuto$^{(1)}{^{\ddagger}}$ \\  
Gudmar Thorleifsson$^{(2)}$\hspace{-.16cm} 
\footnote{\tt thorleif@decode.is \hspace{1.03cm}  
{$^{\S}$}alex@physics.syr.edu}  
\, and \, Alex Travesset$^{(1)}{^{\S}}$\\ \\  
$^1$Physics Department, Syracuse University,\\  
Syracuse, NY 13244-1130, USA  \\   
$^2$ deCODE Genetics, Lynghalsi 1, IS-110 \\  
Reykjavik, Iceland   
}   

\date{}  
  
\maketitle  
  
\begin{abstract}  
  
We present an analysis of extensive large-scale Monte Carlo  
simulations of self-avoiding {\em fixed-connectivity} membranes   
for sizes (number of faces) ranging from 512 to 17672 (triangular) 
plaquettes. Self-avoidance is implemented via {\em impenetrable}  
plaquettes. We simulate the impenetrable plaquette model in both three and four bulk    
dimensions. In both cases we find the membrane to be {\em flat} for all  
temperatures: the size exponent in three dimensions is 
 $\nu=0.95(5)$ (Hausdorff dimension $d_H=2.1(1)$). The single flat phase appears, furthermore, to be  
equivalent to the large bending rigidity phase of {\em  
non-self-avoiding} fixed-connectivity membranes {--} the roughness exponent in three 
dimensions is $\xi=0.63(4)$. This suggests that there is a unique universality class   
for flat fixed-connectivity membranes without attractive interactions.  
 Finally we address some theoretical and experimental   
implications of our work.

\end{abstract}  
  
\vfill\newpage  
  
\pagebreak  
  
\section{Introduction}  
  
There are many different physical systems occurring in nature that we can   
theoretically model as {\em fixed-connectivity} membranes   
\cite{DNJer,D:92,BowTra:00,Wiese:00}. A fixed-connectivity  
membrane is a {\em two-dimensional} mesh, in a $d$-dimensional bulk space,   
whose nodes have a time-independent coordination number. In other  
words the connectivity of the mesh is fixed in time, although there  
may be some spatial variation, provided the resultant density of  
defects (dislocations and disclinations) is small enough to preserve,  
at zero temperature, the quasi-long-range translational order and true long-range bond  
orientational order characteristic of two-dimensional crystals  
\cite{NH:79,Young,Nelson:78}. The statistical mechanics of such  
membranes is a two-dimensional version of the already rich statistical  
mechanics of essentially one-dimensional polymers  
\cite{Jer1,Safran,BowTra:00,Wiese:00}. The physical bulk dimensions  
are $d=2$, corresponding to monolayers or planar crystals, and $d=3$,  
corresponding to membranes. The intrinsic  
importance of this universality class of membranes is enhanced by the  
wealth of examples provided by the natural world.      
These include the spectrin/actin cytoskeleton of mammalian red blood cells  
\cite{Skel:93}, polymerized amphiphilic bilayers and monolayers  
\cite{Fendler2},   
polymerized polymer sheets \cite{Stupp}, graphitic oxides  
\cite{Graphite:91,Graphite:92,Zasa:94},     
metal dichalcogenides \cite{Chianelli} and large sheet molecules of  
glassy $B_2O_3$ \cite{Aziz:85}.    
Experimental measurements of the physical properties of these  
real-world fixed-connectivity membranes can be compared with the results of    
theoretical models and numerical simulations.   
This is an active area of research on the experimental side, with the  
exciting possibility of novel membranes such as two-dimensional  
cross-linked DNA networks.    
A systematic and deep understanding of this field is also  
essential in developing theoretical models of more complex systems,  
such as the complete cell membrane, in which a cytoskeleton  
(a fixed-connectivity membrane) is coupled to an amphiphilic bilayer (itself a  
liquid-like membrane) by protein junctional complexes \cite{Alberts}.       
  
The statistical mechanics of {\em phantom} (self-intersecting)  
fixed-connectivity membranes  
is governed by the competition between the {\em elastic} energy of  
stretching and shearing deformations and the {\em bending} energy of  
shape fluctuations. This competition leads to a scale-dependent  
running of the bending rigidity and elastic moduli of the Hamiltonian  
in the presence of thermal fluctuations,  
with the membrane stiffening under bending and softening under elastic  
deformations as the length scale grows. The upshot is a  
low-temperature ordered phase, characterized by an infinite  
persistence length and a  
spontaneous orientation of the membrane in the bulk space, and a  
high-temperature {\em crumpled} phase with a finite persistence length  
and no preferred orientation in the bulk. The remarkable ordered  
phase, conventionally known by the inapt name {\em flat} phase, is  
one of the key reasons for the excitement in this field. Furthermore  
the global flat and crumpled phases are   
separated by a continuous {\em crumpling} transition associated with  
an ultraviolet-stable fixed point in the renormalization group flow of  
the bending rigidity. The crumpling transition is another exciting  
feature of the physics of phantom fixed-connectivity membranes.  
  
Realistic fixed-connectivity membranes are, however, {\em self-avoiding}  
(SA) in the sense that self-intersections will be energetically  
costly. This self-avoiding, or excluded-volume,                        
interaction arises at the microscopic (nanometer) scale from the  
short range repulsion between any two monomers in the  
membrane. Study of this self-avoiding interaction has proven  
remarkably difficult. Analytical studies (see \cite{BowTra:00} and  
\cite{Wiese:00} for recent reviews) are ambiguous in that there is a  
variety of results for, say, critical exponents. This has made it  
difficult to know the precise nature of the global phase diagram and  
to reliably predict the key critical   
exponents. Numerical results using the balls and springs (BS) model,  
to be discussed in more detail below, clearly indicate that any  
degree of self-avoidance flattens the membrane at all temperatures. In  
other words the crumpled phase is unstable to the inclusion of  
self-avoidance. With its loss goes also the crumpling transition. It  
has been argued, however, that the loss of the crumpled phase is a  
result of the particular {\em discretization} of self-avoidance within  
the BS models, rather than a universal feature of self-avoidance  
itself \cite{AN1:90}. In particular distant-neighbor interactions  
due to the excluded volume of the balls can induce an effective  
bending rigidity for these models even with vanishing bare bending  
rigidity \cite{AN1:90}. Subsequent careful studies of BS  
models support the claim that the crumpled phase is lost (see  
Sec.\ref{SECT__Num_st_SA}). More light can be cast on the problem,  
however, by exploring an alternative discretization of a SA fixed-connectivity  
membrane known as the {\em impenetrable plaquette} (IP) model.  
This will be the focus of the work presented in the rest of the paper.  
Impenetrable plaquette models have the advantage that the  
described membrane is extremely flexible in that individual plaquettes  
are completely free to fold on themselves. As a result they do not  
suffer from induced bending rigidity.   
The organization of the paper is as follows. In the next section we  
provide a quick overview of the BS models for  
completeness. In Sect.\ref{SECT__PModel} we introduce and   
discuss the impenetrable plaquette model studied in this paper. In  
Sect.\ref{SECT__MCsim} we   
describe the details of the algorithms used to numerically simulate   
the model. A complete presentation of our results is given in   
Sect.\ref{SECT__Res} for the cases of bulk dimensions $d=3$ and  
$d=4$.  Finally we summarize in  
Sect.\ref{SECT__conclusions} and discuss the implications of our work  
for the experimentalist and outline some promising future directions.   
  
\section{Self-avoidance: Numerical Studies}\label{SECT__Num_st_SA}  
  
The self-avoidance interaction at long distances (typically of the  
order of a micron) may be modeled by generalizing the Edwards   
model of a self-avoiding polymer \cite{Edwards:65} to objects with  
internal dimension $D$:  
\be\label{Ham_Edu}  
{\cal H}_{SA}=\frac{b}{2}\int d^D {\bf x} d^D {\bf x^{\prime}}   
\delta^d( {\vec r}({\bf x})-{\vec r}(\bf x^{\prime})) \ ,  
\ee  
where $b$ is the dimensionless parameter governing the strength of self-avoidance. 
 
This implementation of self-avoidance captures the underlying  
universal physics for a broad class of microscopic potentials  
describing the interactions between the monomers of the membrane at  
the nanometer scale.   
   
The most direct discretization of fixed-connectivity membranes including  
self-avoidance is a network of $N$ monomers arranged in a   
triangular array. Nearest neighbors interact with a potential \cite{KKN:86}  
\be\label{tether_pot}  
V_{NN}({\vec r})=\left\{ \begin{array}{c c}   
          0 & \mbox{for $\vert {\vec r} \vert < \rho$ } \\  
	  \infty & \mbox{for $\vert {\vec r} \vert > \rho $}  
	                 \end{array} \right.  
			 \ ,  
\ee  
or a smoother variation. The tether length $\rho$ is of the order of a few lattice   
spacings. The self-avoidance, or excluded volume, is introduced as a repulsive hard sphere   
potential, now acting between any two monomers in the membrane, instead   
of only nearest neighbors. A typical hard sphere repulsive potential is  
\be\label{exc_potential}  
V_{Exc}({\vec r})=\left\{ \begin{array}{c c}   
          \infty & \mbox{for $\vert {\vec r} \vert < \sigma $} \\  
	  0 & \mbox{ for $\vert {\vec r} \vert  > \sigma $ }  
	                 \end{array} \right.  
			 \ ,  
\ee  
where $\sigma$ is the range of the potential, and $\sigma <  
\rho$. Once again some   
smoother versions of this potential, continuous at $\vert  
{\vec r} \vert = \sigma$, have also been adopted.   
This model may be pictured as a network of springs, defined by the   
nearest-neighbor potential Eq.\ref{tether_pot}, with self-avoidance  
enforced by the finite radius $\sigma$ of the balls  
(Eq.\ref{exc_potential}) and we will often refer to it as the BS  
(balls and springs) model.   
  
There is conclusive evidence that the phase diagram of the  
three-dimensional BS model consists of a single flat phase {--} the crumpled phase  
being lost entirely [19,22-28].  
A detailed review may be found in \cite{BowTra:00}.  
  
Simulations for bulk dimension above three have also been performed  
\cite{Grest:91,BP:94} and provide clear evidence    
that the membrane remains flat in three and four dimensions and  
undergoes a crumpling transition for bulk dimension $d \geq 5$,  
implying that the lowest dimension, $d_c$, in which a crumpled phase  
exists  is five ($d_c = 5$).   
  
It has been argued \cite{AN1:90} that the absence of a crumpled phase    
is due to induced bending rigidity from distant (next-to-nearest neighbor)  
excluded volume effects. This may place the self-avoiding membrane  
in the flat phase of the phantom membrane phase diagram   
(see \cite{BowTra:00} and references therein), in which case one  
expects the flat phase of the self-avoiding and phantom membranes to  
be equivalent. If this is indeed the case, sufficiently weak excluded volume  
interactions should induce a bending rigidity below the critical crumpling   
transition coupling, leading to a  
crumpled self-avoiding phase. No conclusive evidence for this   
self-avoiding crumpled phase has been reported.  
This suggests that flatness is an inevitable consequence of  
self-avoidance combined with the fixed-connectivity (integrity) of the  
membrane. This picture would be much more convincing if it could be  
shown to apply to other models of self-avoiding fixed-connectivity membranes.  
  
An alternative model of self-avoidance is provided by the    
impenetrable plaquette model (IP), first simulated in  
\cite{B:91,BR:92}. These authors found a size exponent  
corresponding to a crumpled phase and compatible    
with the standard Flory estimate. A subsequent   
simulation \cite{KG:93} of a variation of the IP model, contradicted  
the results of \cite{B:91,BR:92} by claiming a flat phase, but found a  
fractal size exponent, from the finite-size scaling of the    
radius of gyration, of $\nu=0.87$ (no errors quoted). This corresponds  
to a Hausdorff dimension $d_H=2.3$, which is neither Flory nor flat. In  
contrast, the analysis of the orientationally-averaged structure  
function in \cite{KG:93} gives a different    
fractal value $\nu=0.75 (d_H=2.67)$ (no errors quoted).   
In short the status of the phase diagram for the IP model is  
currently murky, in contrast to the clear picture   
emerging from the BS models outlined earlier.  
  
Motivated by this unsatisfactory state of affairs we turn now to an   
analysis of our particular realization of the IP model.

\section{The Impenetrable Plaquette Model}\label{SECT__PModel}  
  
\subsection{GENERAL CASE}  
  
The general discretized energy for a flexible phantom triangular  
fixed-connectivity membrane with bending rigidity $\kappa$ is given by   
\cite{SN:88,BowTra:00}  
\be\label{Gen_action}  
{\cal H}_{\rm ph}=\frac{\epsilon}{2}\sum_{<a,b>} \Bigl( \vert {\vec r}_a-{\vec r}_b  
\vert - 1 \Bigr)^2 +  
\kappa \sum_{<\alpha,\beta>}(1-{\vec n}_{\alpha} \cdot {\vec n}_{\beta})  \ ,  
\ee  
where the elastic term is summed over distinct nearest-neighbor pairs  
of monomers $a$ and $b$ and the bending term is summed over all pairs  
of adjacent triangular plaquettes $\alpha$ and $\beta$. In the free  
energy $F_{\rm ph}$ the parameter $\epsilon$ is a  
discrete version of the elastic moduli (the Lam\'e coefficients) of the  
continuum elastic theory \cite{SN:88}, ${\vec n}_{\gamma}$ is the unit  
normal to the plaquette $\gamma$ and the mean lattice spacing has been  
rescaled to unity without loss of generality. For SA membranes $F_{\rm  
ph}$ must be supplemented by a discretized version of the   
Edwards self-avoidance term introduced in Eq.\ref{Ham_Edu}.   
  
Without self-avoidance the phantom membrane is crumpled for all  
temperatures above the critical crumpling temperature {--} the  
microscopic bending rigidity is scale-dependent and driven to zero at  
long wavelength by thermal fluctuations. In this regime only the  
elastic term is relevant in Eq.\ref{Gen_action}. Under the rescaling  
of the bulk coordinate ${\vec r} \rightarrow {\vec r}/\sqrt{\beta}$  
we find    
\be\label{Gen_action_hight}  
\beta {\cal H}=\frac{\epsilon}{2}\sum_{<a,b>} \left( {\vert {\vec r}_a-{\vec r}_b \vert}^2   
- 2\sqrt{\beta}\vert {\vec r}_a-{\vec r}_b\vert+\mbox{const.}\right) \ ,  
\ee  
which reduces to the pure Gaussian model in the high-temperature limit  
$\beta \rightarrow 0$.  
  
Suppose we can establish that self-avoidance flattens the Gaussian  
model, which corresponds to the elastic term in Eq.\ref{Gen_action}  
with vanishing mean lattice spacing. Then clearly the SA model  
 Eq.\ref{Gen_action}, together with self-avoidance, will also be flat  
at all temperatures.     
  
\subsection{DISCRETIZATION}\label{sub_SECT__model}  
  
The continuum Edwards model Eq.\ref{Ham_Edu}, for  membranes ($D=2$),    
is then  
\be\label{Ham_Edw_2D}  
{\cal H}=\int d^2 {\bf x} \parp_{\alpha} {\vec r} \parp^{\alpha} {\vec r}  
+\frac{b}{2}\int d^2 {\bf x} d^2 {\bf x^{\prime}}   
\delta^d( {\vec r}({\bf x})-{\vec r}(\bf x^{\prime}) ) \ ,  
\ee  
where $d$ is the bulk dimension.  
  
Let's apply this  to piecewise flat surfaces defined by $N$ vertices 
$({\vec r_a)}_{1\leq a \leq N}$, with all vertices, except those at 
the boundary, being six-coordinated. Any point on the surface   
is within a triangle defined by its three vertices   
$\{{\vec r}_a,{\vec r}_b,{\vec r}_c\}$, and may be parametrized as  
\be\label{surface}  
{\vec r}(\alpha,\beta)=\alpha {\vec r}_a+\beta {\vec r}_b+  
(1-\alpha-\beta){\vec r}_c \ ,  
\ee  
where $\alpha\geq 0$, $\beta \geq 0$ and $\alpha+\beta \leq 1$.  
  
The first term in Eq.\ref{Ham_Edw_2D} is easily discretized as the  
Gaussian term previously discussed   
\be\label{Gauss_term}  
\int d^2 {\bf x} \parp_{\alpha} {\vec r} \parp^{\alpha} {\vec r}=  
\frac{1}{2}\sum_{<a,b>} ({\vec r}_a-{\vec r}_b)\cdot ({\vec r}_a-{\vec r}_b)  
\ ,  
\ee  
whereas for the self-avoiding interaction we have  
\bea\label{SA_term}  
\int d^2 {\bf x} d^2 {\bf x'} \delta^d(\vec{r}({\bf x})-\vec{r}({\bf x}'))  
&=& \int d^2 {\bf x} d^2 {\bf x'} \prod_{\mu=1}^{d}  
\delta ({r^{\mu}}({\bf x})-{r^{\mu}}({\bf x}'))   
\\\nonumber  
&=&\sum_{i\neq j} \int d\alpha_i d\beta_i d\alpha_j d\beta_j  
\prod_{\mu=1}^{d} \delta ({r^{\mu}}({\bf x})-{r^{\mu}}({\bf x}'))\ ,  
\eea  
where the last sum runs over all distinct pairs of    
triangles $i,j$. Each term here can be explicitly evaluated   
for any  given pair of triangles. Non self-intersecting pairs do not  
contribute to the sum while for self-intersecting triangles the result  
depends on the dimensionality and the precise type of self-intersection:   
   
\begin{figure}[t]  
\centerline {\epsfig{file=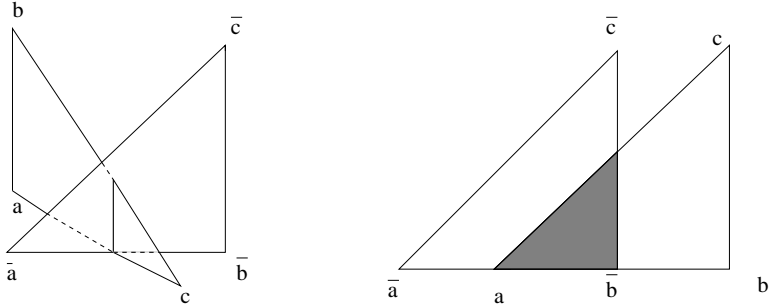,width=4in}}  
\caption{Two intersecting triangles in three dimensions. In the first  
case the self-avoiding energy is finite, while in the second it is  
infinite.}   
\label{fig__twotriang}  
\end{figure}  
  
\begin{itemize}  
\item {$d=3$:}  
Triangles may self-intersect in two different ways, as shown in   
Fig.\ref{fig__twotriang}. In the first case the self-intersection is  
a line whereas in the second it is a two-dimensional  
region. One-dimensional self-intersections have a finite energy given by   
\be\label{finite_energy}  
\int d\alpha d \beta d \vet{\alpha} d \vet{\beta} \hspace{0.1cm}  
\delta^{(3)}\Bigl({\vec r}(\alpha,\beta)-{\vec  
r}(\vet{\alpha},\vet{\beta})\Bigr)   
\propto s \ ,  
\ee  
where $s$ is the intersection length.  
Two-dimensional self-intersections, however, have infinite energies  
(see Fig.\ref{fig__twotriang}).   
  
\item{$d=4$:}  
The integral is finite if triangles self-intersect in a point but  
infinite if they intersect in a line or a surface.  
  
\item{$d \geq 5$}  
The integral is infinite whenever there is self-intersection.   
  
\end{itemize}  
  
The model we treat is an open triangular network containing $N$ vertices   
with free boundary conditions. The energy is  
\be\label{Ham_dis}  
{\cal H}=\frac{1}{2}\sum_{<ab>}   
{\vert {\vec r}_a-{\vec r}_b \vert}^2  \ ,  
\ee  
provided the triangles defining the surface do not intersect  
anywhere. If two triangles do intersect the configuration is strictly  
forbidden (its energy is infinite).   
This discretization corresponds to tuning the   
self-avoiding coupling in Eq.\ref{Ham_Edw_2D} to infinity ($b=\infty$).   
It may be worthwhile  to consider a weaker version of self-avoidance,  
corresponding to a finite value of $b$, in which some degree of  
self-intersection is permissible. This will not be treated in this  
paper since the computational cost rises significantly and we do not  
expect the critical behavior to be sensitive to the precise value of  
$b$.      
  
\begin{figure}[htb]  
\centerline{\epsfig{figure=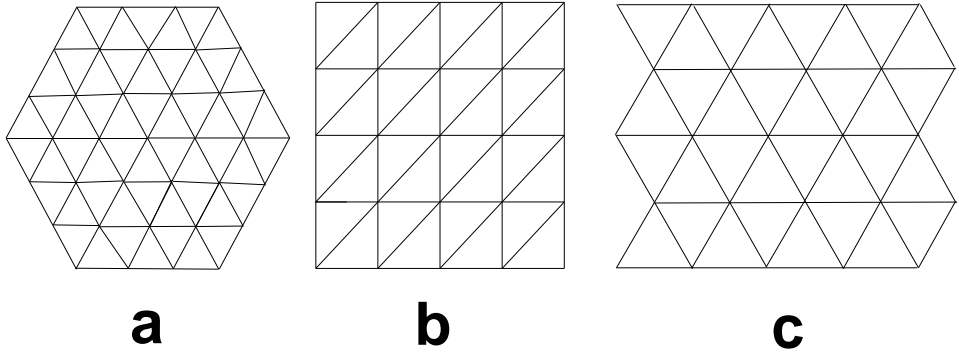,width=5in}}  
\caption{The three lattice geometries adopted in our membrane  
simulations, where $L=3$ for {\bf a} and $L=5$ for {\bf b} and {\bf c}.}  
\label{fig__network}  
\end{figure}

We adopted three distinct geometries, each depicted in Fig.\ref{fig__network},  
in our simulations. The 3d simulations employed mostly the $c$  
geometry, but the other two geometries were explored as a test of  
universality. The 4d simulations used the hexagonal geometry $a$.   
  
The square geometries $b$ and $c$, with $L$ vertices on a side, have a  
total number of nodes $N=L^2$.    
The hexagonal geometry $a$, with $L$ links on each of the six  
sides, has a total number of nodes   
\be\label{hexagon_geom}  
N=3L^2+3L+1 \ .  
\ee  
  
\subsection{OBSERVABLES and MEASUREMENTS}  
  
A comprehensive understanding of the {\em shape} of a fluctuating 
membrane may be obtained from a careful analysis of the shape 
tensor\footnote{The shape tensor is obtained from the standard moment 
of inertia tensor by subtracting out the isotropic contribution and is 
sometimes also mistakenly called the moment of inertia tensor in the 
literature.} \cite{AN1:90,KK:93,BCFTA:96},   
\be\label{Inert_ten}  
S_{\alpha \beta}=\frac{1}{2N^2}\sum_{k=1}^N\sum_{l=1}^N(r_{\alpha k}-r_{\alpha l}) (r_{\beta k}-r_{\beta l})=\frac{1}{N}\sum_{k=1}^N r_{\alpha k} r_{\beta k} -\frac{1}{N^2} \sum_{k=1}^N r_{\alpha k} \sum_{l=1}^N r_{\beta l} \ ,  
\ee   
where the indices $k,l$ run over the lattice.  
The trace of $S_{\alpha \beta}$ is the squared radius of gyration 
$R_G^2$ 
\begin{equation} 
R_G^2={\rm Tr} \, S_{\alpha \beta} \, . 
\end{equation} 
 The growth of the radius of gyration with internal system 
size $L$ determines the Hausdorff dimension $d_H$, or equivalently, 
the size exponent $\nu=2/d_H$, via $R_G \sim L^{\nu}$. 
Diagonalizing the (symmetric) shape tensor and performing a 
statistical average over independent configurations defines $d$ shape 
exponents $\beta_{\alpha}$ via the finite-size scaling  of its 
eigenvalues $\lambda_{\alpha}$   
\be\label{Eigen_scl}  
\langle \lambda_{\alpha} \rangle \sim N^{\beta_{\alpha}} \ ,  
\ee  
with $\alpha=1,\cdots,d$. In a crumpled (isotropic) phase all $d$ 
exponents are equal and given by $\nu$ (i.e. $\beta_{\alpha}=\nu$ 
for all $\alpha$).  
In a flat phase, on the other hand, $\beta_{\alpha}=1$ for the two 
in-plane eigen-directions (say $\alpha=1,2$) and the remaining shape 
exponents describe the root-mean-square height (${\vec h}$) 
fluctuations. For $d=3$, to be specific, $\beta_3$ is the standard   
roughness exponent $\zeta$ ($<h^{2}> \sim L^{2\zeta}$) 
\cite{AN1:90,AN2:90}.     
  
Further important information on the conformation of the  
membrane is provided by the structure function, defined as   
\be\label{Struct_fun}  
S({\vec q} \,)=\frac{1}{N^2}\sum_{a,b}  
\Bigl{\langle} e^{i{\vec q} \cdot ({\vec r}_a-{\vec r}_b)} \Bigr{\rangle} \ ,  
\ee  
where ${\vec q}$ is an arbitrary wavevector.   
In particular the scaling properties of the structure function for 
wavevectors parallel to the eigenvectors of the shape tensor will 
allow us to determine the key shape exponents of the model.    
  
\section{Simulation Methods}\label{SECT__MCsim}  
  
Simulations of self-avoiding membranes are hampered both by long   
autocorrelation times in updating the embedding  
coordinates and by the non-locality of the self-avoidance   
constraint. These two factors combined have effectively  
prevented simulations of large enough SA fixed-connectivity  
membranes for a reliable determination of their scaling 
properties. There has been some recent progress though in   
overcoming both these problems.  
  
\vspace{10pt}  
Improved methods for updating the embedding $\{ {\vec r}({\bf x}) \}$ have been applied   
in simulations of a {\it non-SA} (phantom) fixed-connectivity membranes.  
In \cite{TF:98} three different methods were compared:  
\begin{itemize}  
 
 \item  
   A standard Metropolis updating scheme.  
 
 \item  
   Hybrid over-relaxation:  make a quadratic approximation  
   to the action\footnote{Note that over-relaxation is exact for 
   vanishing bending rigidity since the action is quadratic.},  
   then apply over-relaxation followed by a  
   Metropolis accept/reject test.  
  
 \item  
   Unigrid algorithm: update the membrane recursively on all   
   length scales by dividing the lattice into sub-lattices of  
   different sizes and apply a Metropolis algorithm to a collective  
   update of those parts.  
\end{itemize}  
 
\begin{figure}[t]  
\centerline{\epsfig{figure=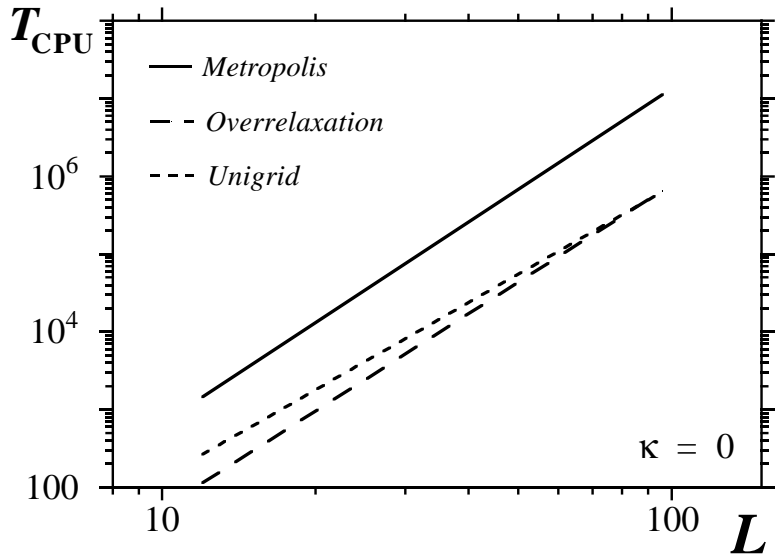,width=4in}}  
 \caption{The performance of three different methods  
  for updating the embedding of an isotropic  
  phantom fixed-connectivity membrane.}  
  \label{fig__phantom}  
\end{figure}

The performance of each of these methods {--} the CPU-cost per  
independent configuration ($T_{\rm CPU}$) {--} is compared in  
Fig.\ref{fig__phantom}. This figure is based on the values   
presented in \cite{TF:98}.    
For the membrane sizes typically used in Monte Carlo  
simulations, $L < 100$, both the hybrid over-relaxation   
and the unigrid algorithm reduced the cost ten-fold  
compared to a simple Metropolis algorithm.  Only the  
unigrid algorithm, however, reduces the dynamical exponent  
$z$, which measures the volume scaling of the CPU-cost ($T_{\rm 
CPU} \sim L^z$) from $z=4$ to $z\approx 3.8$.

\vspace{8pt}  
  
The superior over-relaxation and unigrid algorithms are both easily adapted to simulations   
of SA fixed-connectivity membranes.  The real CPU-cost in self-avoiding 
simulations, on the other hand, is the non-locality of the updating.    
Implementing SA as impenetrable plaquettes requires a   
check verifying that the proposed update  
of the embedding does not lead to intersecting triangles.   
In addition to being non-local, pairwise intersection-checking is 
very time-consuming. It is, however, possible to reduce this CPU-cost 
by a clever implementation of the SA check.  We have compared three 
such implementations:   
\begin{itemize}  
  \item[{\bf A}]  
  A comprehensive check; explicitly verifying that no   
  updated triangle intersects any other triangle in the lattice   
  by comparing all relevant pairs of triangles.  
  \item[{\bf B}]  
  Inscribe each triangle in a {\it minimal} sphere in the embedding  
  space.  This has the advantage that pairs of triangles whose  
  spheres do not overlap are not compared, and overlapping   
  are quickly identified.   
  \item[{\bf C}]  
  Inscribe regions (sub-lattices) of the membrane  
  of different sizes in a minimal sphere and  
  apply method {\bf B} {\it recursively} to regions of decreasing size.  
  In this way large portions of the membrane are quickly  
  eliminated from the checking procedure.  
  \end{itemize}  
For simplicity  we use these methods in combination with a hybrid 
over-relaxation algorithm as it updates the membrane locally. 
Although it is not a priori clear that the unigrid algorithm, which 
performs a non-local update, should perform any worse than  
the over-relaxation applied to SA-membranes, it would be more complicated  
to implement the above methods.  
  
In Fig.4 we compare the CPU-times for one sweep of hybrid over-relaxation  
for each of these methods.  Methods {\bf B} and {\bf C} reduce the 
overall CPU-time substantially and, more importantly, both   
improve the volume scaling of the CPU-cost:  
from $t_{\rm sw} \sim L^{4.4}$ for method {\bf A} to $t_{\rm sw} \sim 
L^{2.4}$ for method {\bf C}.  
  
\begin{figure}[t]  
 \centerline{\epsfig{figure=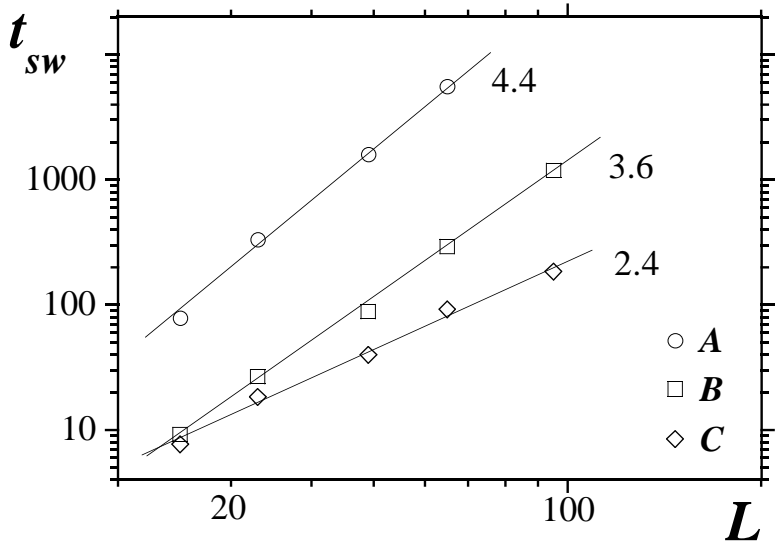,width=4in}}  
 \caption{The CPU-times for updating a SA fixed-connectivity  
  membrane ($d=3$) using hybrid over-relaxation with three different methods   
  of checking for self-intersection. For each method an estimate of  
  the computational-cost exponent is included.}  
  \label{fig__SA__CPU}  
\end{figure}  
  
The simulations were done on several different machines; Pentium II  
(250 MHz), IBM SP2 (160 MHz) and a DEC ALPHA workstation cluster.   
The total computational cost of the three-dimensional simulation was  
equivalent to 150,000 SP2 hours while that for the four-dimensional  
simulation was 60,000 SP2 hours.  
The total number of sweeps depends on the volume and   
configurations are stored after a certain fixed number of sweeps.   
The total number of independent configurations for each volume and  
dimension can be read off from Tables \ref{tab__AT3d}\footnote{Note 
that the autocorrelation times in this table do not give the correct 
finite-size scaling exponent $z$ because the larger volumes were 
simulated with a more effective algorithm than the smaller volumes.}  
  and \ref{tab__ATvol}.  
  
We next address the quality of our data. Given an observable $O$ we 
measure, after thermalization, the autocorrelation time $\tau_O$     
\be\label{AT_time}  
{\tau}_O(T)=\frac{1}{2}+\sum_{t=1}^{T} \rho_O(t) \ ,  
\ee  
where the normalized autocorrelation function $\rho_O(t)$ is given by  
\be\label{fun_AT}  
\rho_O(t)=\frac{  
(\frac{1}{M-t}\sum_{i=1}^{M-t} O_{i} O_{i+t}-\langle O \rangle ^2)}{  
(\langle O^2 \rangle-\langle O \rangle^2)} \ . 
\ee  
and $T$ is a cutoff generally taken to be the largest  
value for which $\rho$ does not become negative \cite{Sokal}.   
  
\begin{table}[tb]  
\centerline{  
\begin{tabular}{|r|r|c||r|r|c|}  
\multicolumn{1}{c}{$N$} & \multicolumn{1}{c}{$\tau_{R_G}$} 
 &\multicolumn{1}{c}{IND CONF}& \multicolumn{1}{c}{$N$} & 
\multicolumn{1}{c}{$\tau_{R_G}$} & \multicolumn{1}{c}{IND CONF} \\\hline  
289   & $2000$ & $2500$ &  2401  & $20000$ & $150$    \\\hline   
625   & $7000$ & $1300$& 4225  & $40000$ & $80$  \\\hline  
1089  & $12000$ & $350$ & 9025  & $71000$ & $30$    \\\hline  
\end{tabular}}  
\caption{Number of independent configurations (IND CONF) for each 
volume for $d=3$.}  
\label{tab__AT3d}  
\end{table}  
\begin{table}[tb]  
\centerline{  
\begin{tabular}{|r|r|r|c|c|}  
\multicolumn{1}{c}{$N$} & \multicolumn{1}{c}{MC ($\times 10^5$)}   
& \multicolumn{1}{c}{TH($\times 10^4$)}   
& \multicolumn{1}{c}{$\tau_{R_G}$}   
& \multicolumn{1}{c}{IND CONF}   
\\\hline  
 61  & $5.5 $ & $ 5  $  &  $  8.2(6)$&$67073  $  \\\hline  
127  & $8.0 $ & $10  $  &  $ 26.0(9)$&$30769  $  \\\hline  
217  & $8.0 $ & $10  $  &  $ 80(10) $&$10000 $  \\\hline  
331  & $13.0$ & $30  $  &  $210(35) $&$6190  $ \\\hline  
469  & $12.0$ & $20  $  &  $400(40) $&$3000   $ \\\hline  
817  & $12.0$ & $32  $  &  $1000(100)$&$1200  $   \\\hline  
1261 & $16.0$ & $40  $  &  $3400(700)$&$470   $   \\\hline  
\end{tabular}}  
\caption{Number of Monte Carlo sweeps performed for each volume for  
bulk dimension $d=4$, where a Monte Carlo sweep is defined as a 
combined Metropolis and over-relaxation sweep. The number of 
thermalization sweeps is also listed, together with the 
autocorrelation times of the shape tensor and the net number of 
independent configurations.}   
\label{tab__ATvol}  
\end{table}  
    
\begin{figure}[htb]  
\centerline{\epsfig{figure=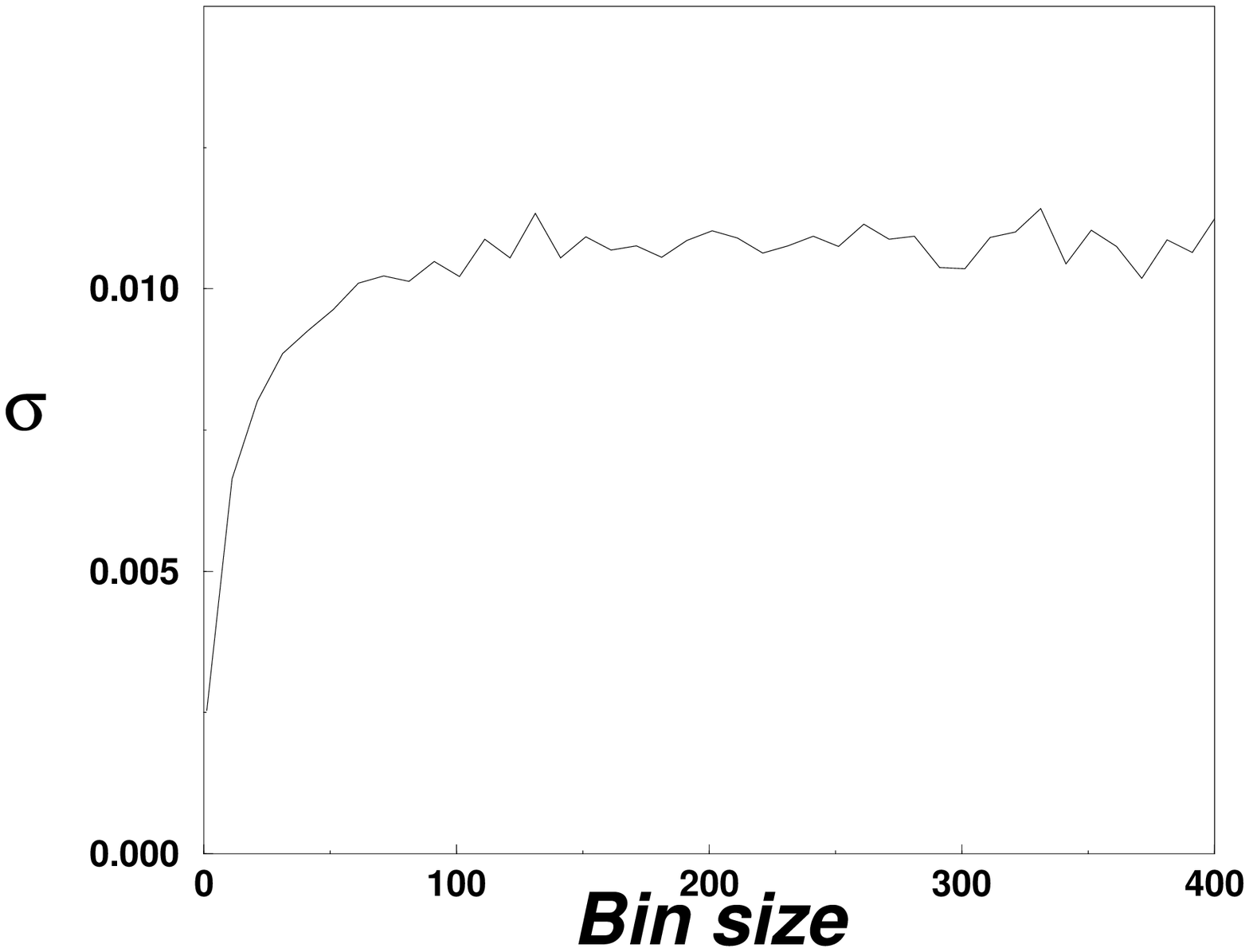,width=2.8in}  
\epsfig{figure=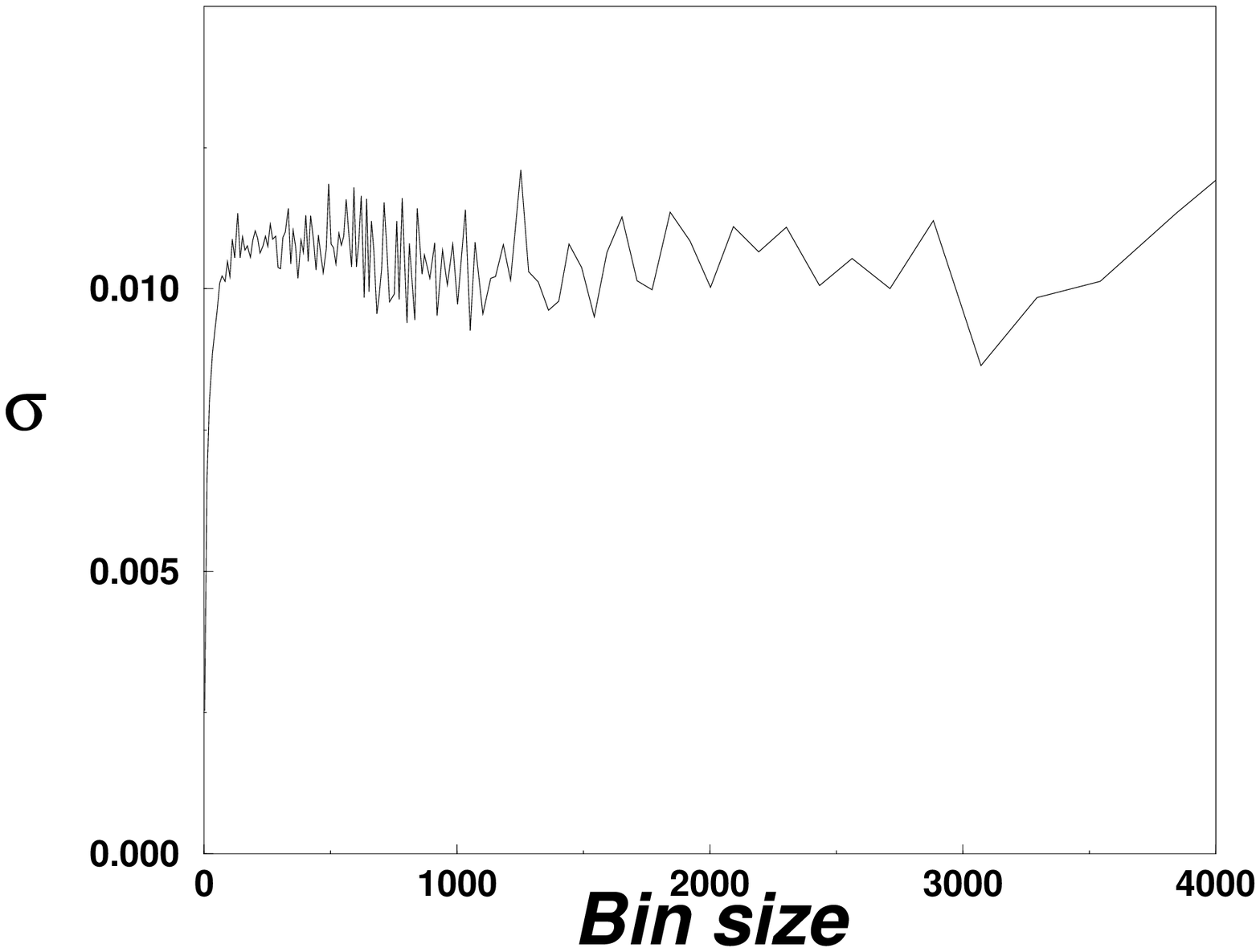,width=2.8in}}  
\caption{Plot of a typical error bar as a function of the bin size.   
The left figure magnifies the small size bins of the right one.  
This particular plot correspond to the observable $S_{22}$ at volume   
$N=331$ ($d=4$). Measurements were taken every 20 sweeps.}  
\label{fig__error_bin}  
\end{figure}  
  
In Tables \ref{tab__AT3d} ($d=3$) and \ref{tab__ATvol}($d=4$) 
 we list autocorrelation times for the shape tensor,  
one of the slowest observables of the membrane.  
There is a striking dependence on volume. 
  
A cross-check of statistical independence may be performed    
by studying the error bars as a function of bin size. It is well known   
that the error bar for any observable $O$ scales as a function of the   
bin size as  
\be\label{error_bar_sc}  
\sigma_O(n_b)/\sigma_O(1) \sim \sqrt{(2\tau_O+1)}  
\ee  
where the bin size $n_b$ must be big enough for bins to be   
statistically independent. To further ensure the statistical 
independence of our data, we verified this dependence in all our 
measurements. As a concrete example, we plot in 
Fig.\ref{fig__error_bin} the error bar for $S_{22}$ as a function of 
the bin size. A clear plateau is observed corresponding to an error 
bar $\sigma=0.011$. We also find $\sigma(1)=0.0025$.    
>From Eq.\ref{error_bar_sc} we get ${\tau_{eff}=9.2}$. Since   
measurements were taken every 20 sweeps, we have   
${\tau_{R_G}=20}$ and ${\tau_{eff}=184}$, in good agreement with the   
result given in Table \ref{tab__ATvol}.   
  
In summary we believe that we have done the requisite checks to ensure  
the statistical independence of our data and we move on to a  
discussion of results.  
  
\section{Measurement of Observables}\label{SECT__Res}  
  
\subsection{BULK DIMENSION $d=3$}\label{Sub_SECT__d3}

\begin{figure}[htb]  
\centerline {\epsfig{file=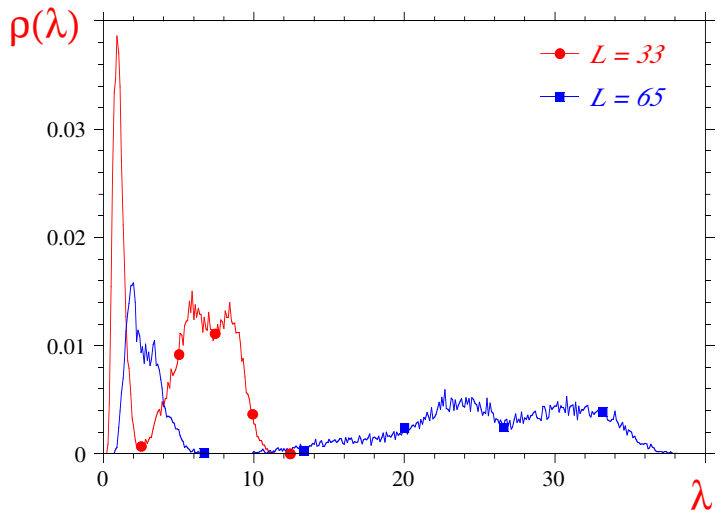,width=12cm}}  
\caption{Distribution of the eigenvalues of the shape  
tensor for volumes $N=1089$ and $N=4225$ ($d=3$).}  
\label{fig__no3d__crumpled}  
\end{figure}

The first observables we analyze are the eigenvalues of the   
shape tensor Eq.\ref{Inert_ten}.  
In Fig.\ref{fig__no3d__crumpled} we plot the distribution of the shape  
tensor eigenvalues for $L=33$ ($N=1089$) and $L=65$ ($N=4225$).   
There is a clear first peak, which we identify   
with height fluctuations, and a second peak which we identify  
with size fluctuations. The slight double well structure of the second  
peak is a reflection of the asymmetry of the adopted geometry,    
as illustrated in Fig.\ref{fig__network}.  
The overall double peak distribution of eigenvalues corresponds to an 
anisotropic surface resulting from rotational symmetry breaking and is 
the first signal of an orientationally ordered ({\em flat}) phase.  
Indeed, the distribution of eigenvalues found here is similar to that 
found in the flat phase of a phantom fixed-connectivity membrane 
\cite{BCFTA:96}.   
 
We next determine the three exponents $\beta_{\alpha}$ from the power 
law scaling given in Eq.\ref{Eigen_scl}. The results are shown in   
Tables \ref{tab__3d_lamb1} and \ref{tab__3d_lamb23}. The column corresponding to Fit1 is a  
direct fit to the power law, including all volumes, whereas Fit2 excludes  
volumes $N < 625$. To investigate the role played by the boundary   
(we have free boundary conditions), we also perform fits excluding   
nodes near the boundary. Excluding nodes beyond $0.9\times L$ from the 
center, we perform a full fit (Fit3) or a fit removing sizes $N < 625$  
(Fit4). We also report similar fits (Fit5 and Fit6)     
excluding nodes beyond $0.8\times L$ from the center.  
It is evident that the membrane does not exhibit strong edge fluctuations. 
    
\begin{table}[tb]  
\centerline{  
\begin{tabular}{|c||l|l||l|l||l|l|}  
\multicolumn{1}{c}{} & \multicolumn{1}{c}{Fit1} &  
\multicolumn{1}{c}{Fit2} & \multicolumn{1}{c}{Fit3} &   
\multicolumn{1}{c}{Fit4} & \multicolumn{1}{c}{Fit5} &  
\multicolumn{1}{c}{Fit6}   
\\\hline  
$\nu$       & $0.945(14)$ & $0.959(13)$  &  $0.938(13)$   
& $0.948(11)$  & $0.918(14)$ & $0.944(13)$ \\\hline  
$\zeta$ & $0.655(22)$ & $0.660(33)$  &  $0.649(19) $   
& $0.655(30)$ & $0.642(16)$ & $0.652(23)$ \\\hline   
\end{tabular}}  
\caption{Results of the various fits to a simple power law, as  
described in the text, for the size exponent $\nu$ and   
the roughness exponent $\zeta$ ($d=3$).}  
\label{tab__3d_lamb1}  
\end{table}  
  
This analysis gives a size exponent $\nu$ near one and   
a roughness exponent $\zeta=0.65(1)$. Our results show a slight  
dependence on the volumes included in the fit, indicating that sub-leading  
corrections are not negligible. We repeated the fits using two 
different parametrizations of the sub-leading corrections. This is shown in  
Table \ref{tab__3d_lamb23}. This improved the quality of the fits  
considerably. The size exponent $\nu$ moves towards one and the  
roughness exponent $\zeta$ decreases slightly.  
   
\begin{table}[tb]  
\centerline{  
\begin{tabular}{|c||c|c||c|c|}  
 \multicolumn{1}{c}{fit}      & \multicolumn{1}{c}{$\nu$} &  
 \multicolumn{1}{c}{$\chi^2$} & \multicolumn{1}{c}{$\zeta$} &   
 \multicolumn{1}{c}{$\chi^2$}   
\\\hline  
$aN^{\beta}+b$   & $0.945(14)$ & $5.7$  &  $0.655(22)$ & $0.994$  \\\hline  
$aN^{\beta}+b\log(N)$ & $0.959(15)$ & $5.3$  &  $0.63(1)$ & $5.9$  \\\hline  
\end{tabular}}  
\caption{Comparison of the size and roughness exponents   
with two distinct functional forms of sub-leading corrections ($d=3$).  
The whole lattice is included in these fits.}  
\label{tab__3d_lamb23}  
\end{table}  
  
\begin{figure}[htb] 
\centerline {\epsfig{file=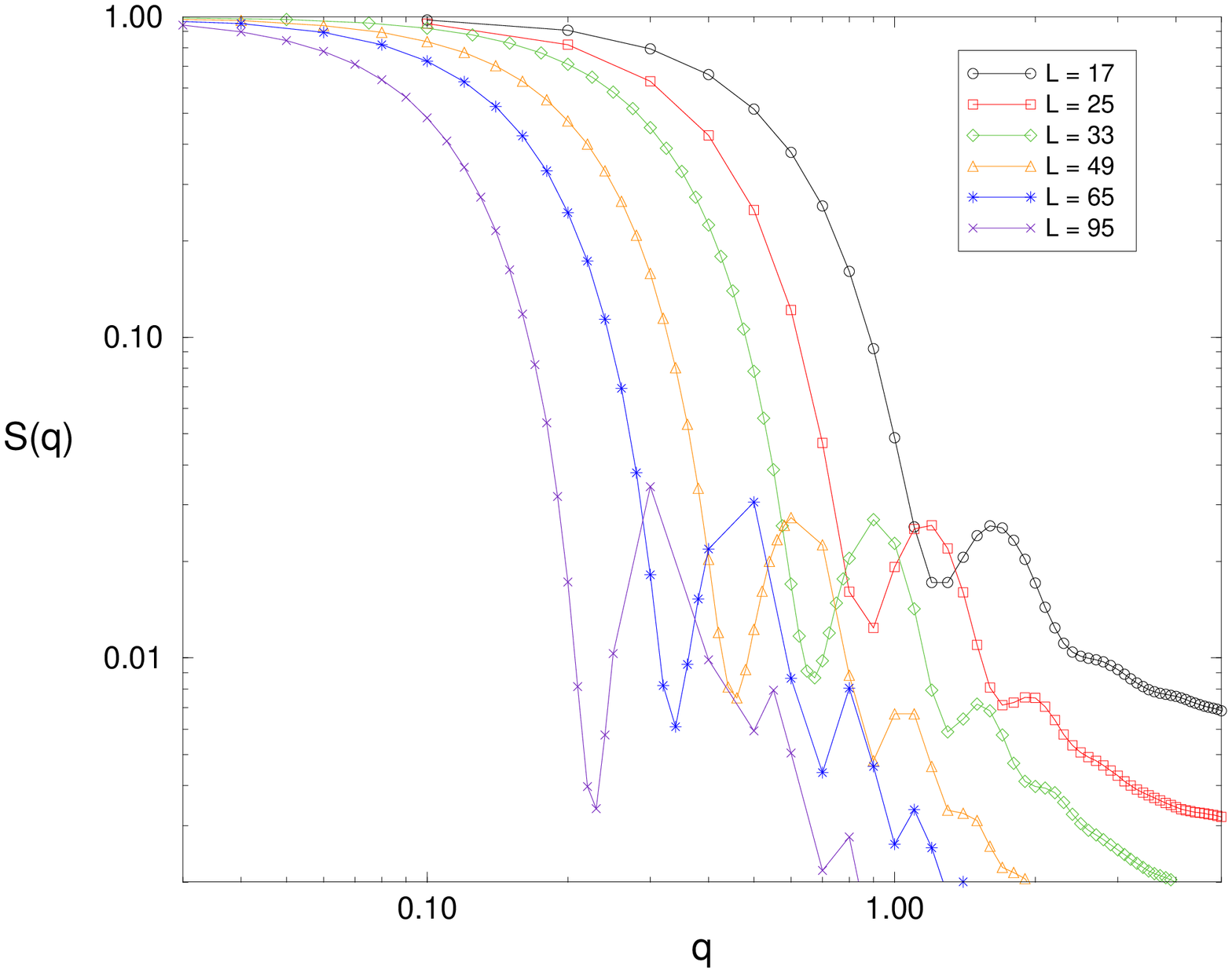,width=8cm,angle=270}}  
\caption{Log-Log plot of the structure functions along the direction of 
the maximum eigenvalue of the shape tensor as a function of the wave 
vector $q$ ($d=3$).}   
\label{fig__allstruct}  
\end{figure}

Further progress requires study of the structure function,   
Eq.\ref{Struct_fun}. In Fig.\ref{fig__allstruct} we plot the  
structure function along the directions of the largest eigenvalue for 
different lattice sizes. For small momentum the structure function is 
monotonically decreasing.  For $q$ sufficiently large, however, a 
series of peaks appear.  
  
\begin{figure}[hp]  
\centerline {\epsfig{file=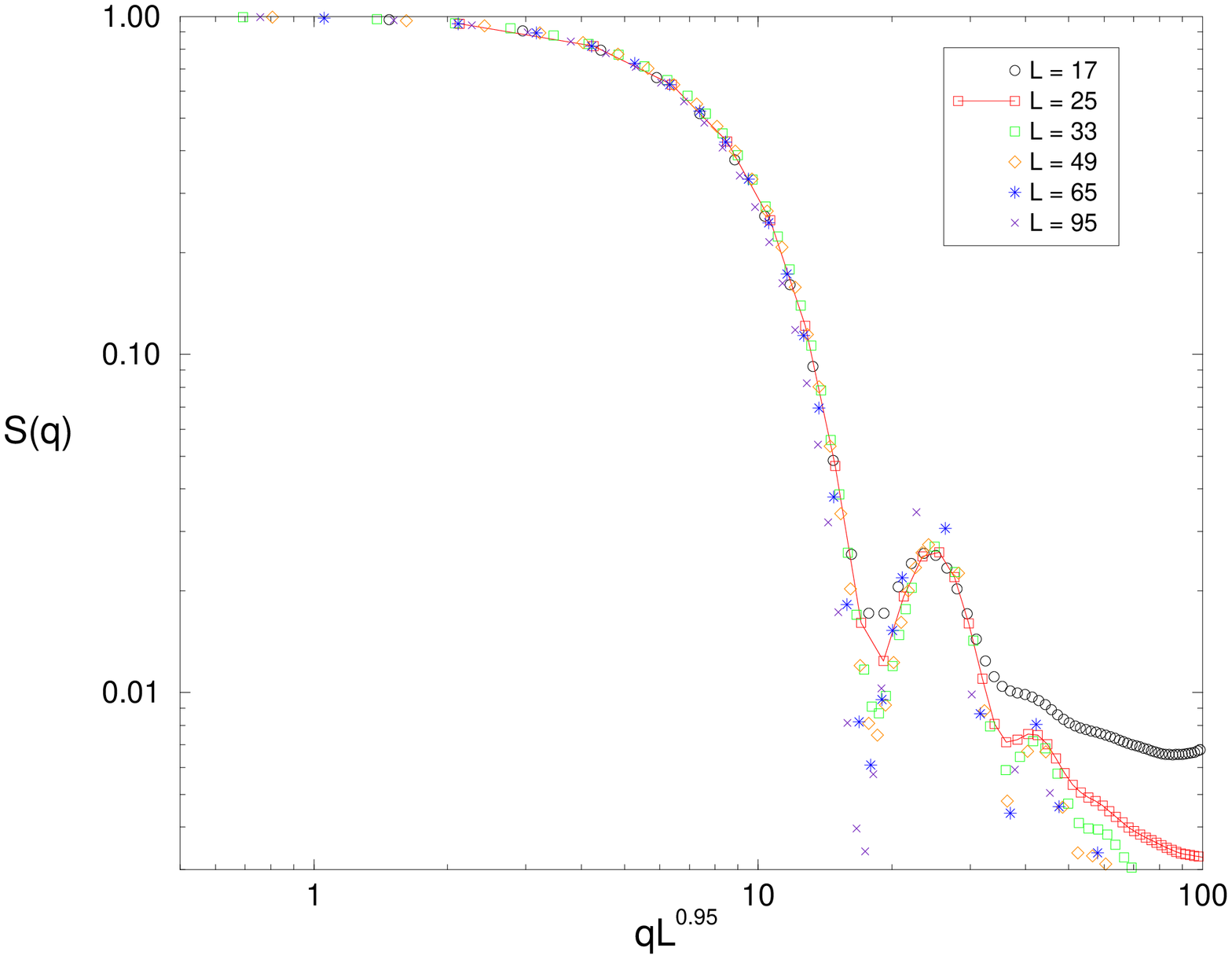,width=8cm,angle=270}}  
\caption{Log-Log plot of the structure function along the direction of 
the maximum eigenvalue of the shape tensor, as a  
function of the scaling variable, for each volume simulated ($d=3$).}  
\label{fig__sq3d1}  
  
\centerline {\epsfig{file=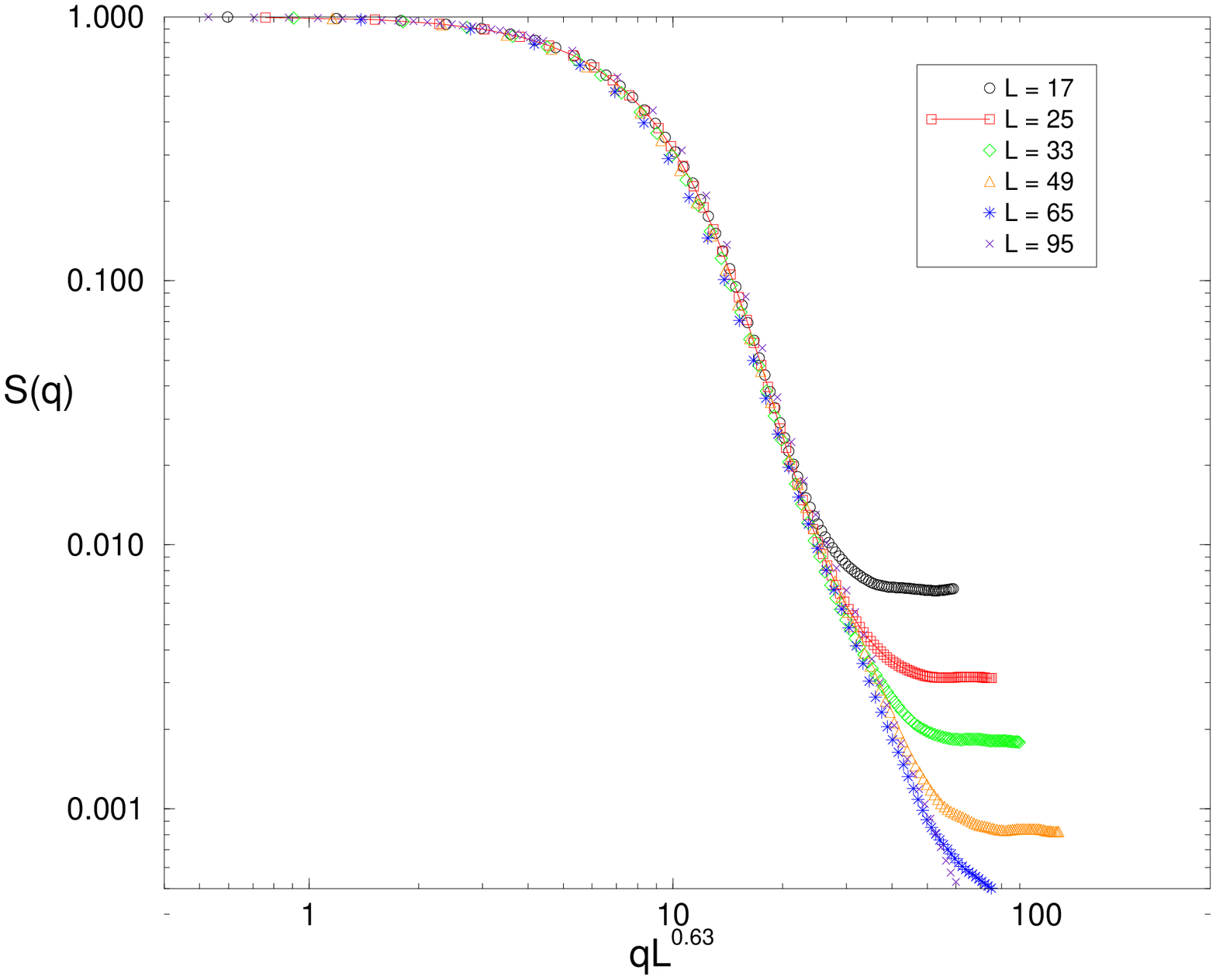,width=8cm,angle=270}}  
\caption{Log-Log plot of the structure function along the direction 
normal to the preferred plane of the membrane, as a function of the 
scaling variable, for each volume simulated ($d=3$).}  
\label{fig__sq3d2}  
\end{figure}  
  
>From the small-$q$ region we can extract the critical exponents $\nu$ and  
$\zeta$. To do so we plot the structure function along the directions 
corresponding to the largest and smallest eigenvalues, respectively, as 
a function of the scaling variables $qL^{\nu}$ and $qL^{\zeta }$.  
Collapse to a single scaling curve is excellent for $\nu=0.95(5)$  
and $\zeta=0.63(4)$ as illustrated in Figs.\ref{fig__sq3d1} ($\nu=0.95$) 
and \ref{fig__sq3d2} ($\zeta=0.63$) 
\footnote{The quoted errors for these scaling exponents are rather 
insensitive to the precise statistical method by which the quality of 
the scaling collapse is assessed.} 
 
\begin{figure}[htb]  
\centerline {\epsfig{file=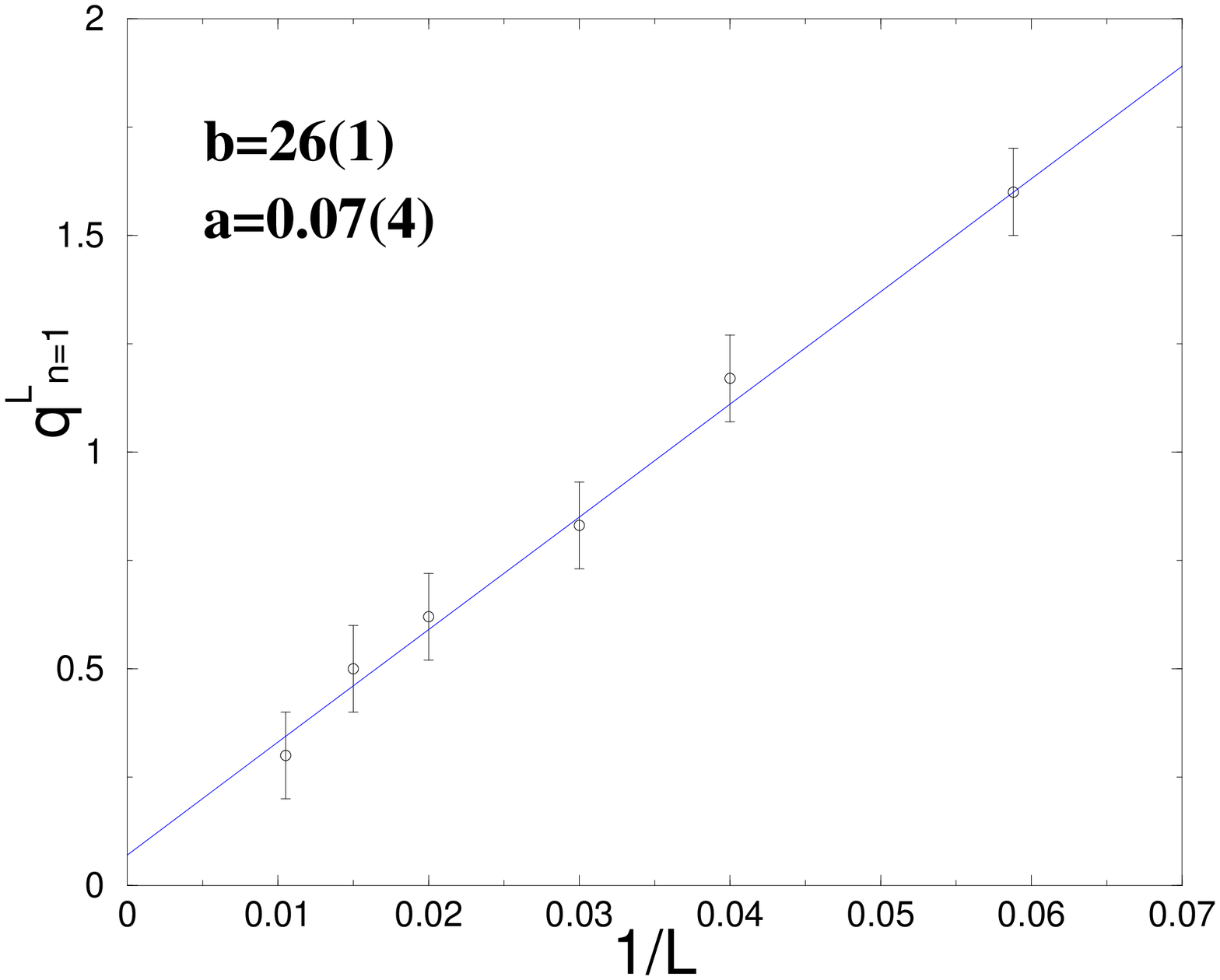 ,width=6.5cm,angle=270}   
\epsfig{file=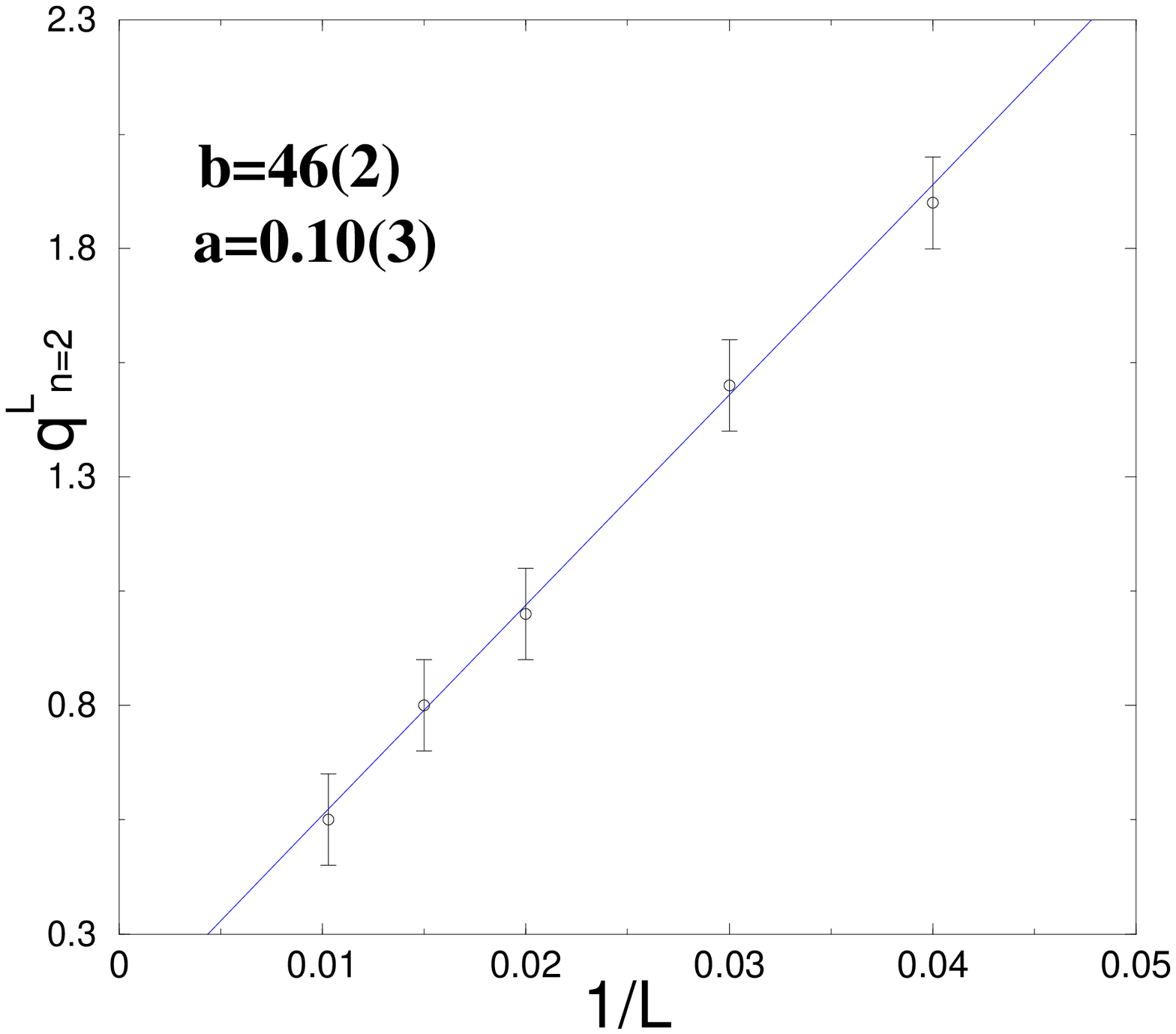 ,width=6.5cm,angle=270} }  
\caption{Fits of the peaks of the structure function to the form  
$q_{n}^{L}=a + b_n/L$ ($d=3$).}  
\label{fig__q__peaks}  
\end{figure}  
  
\begin{figure}[hp]  
\centerline {\epsfig{file=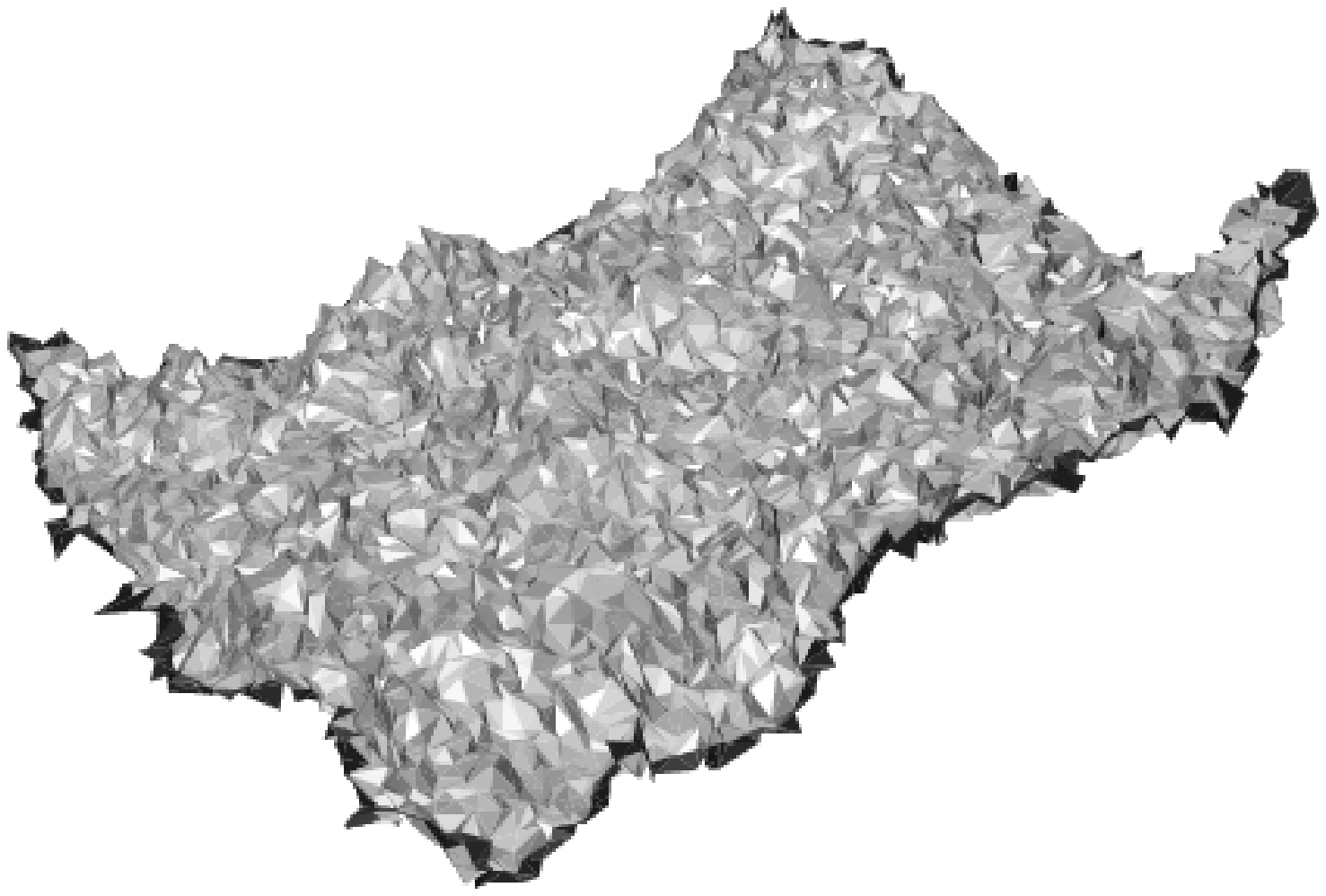,width=8cm,angle=90}}  
\caption{A snapshot of a thermalized {\em self-avoiding} configuration   
for volume $N=9025$. The membrane is flat over long length scales but rough   
on short scales ($d=3$).}  
\label{fig__snapshot}  
\centerline{\epsfig{figure=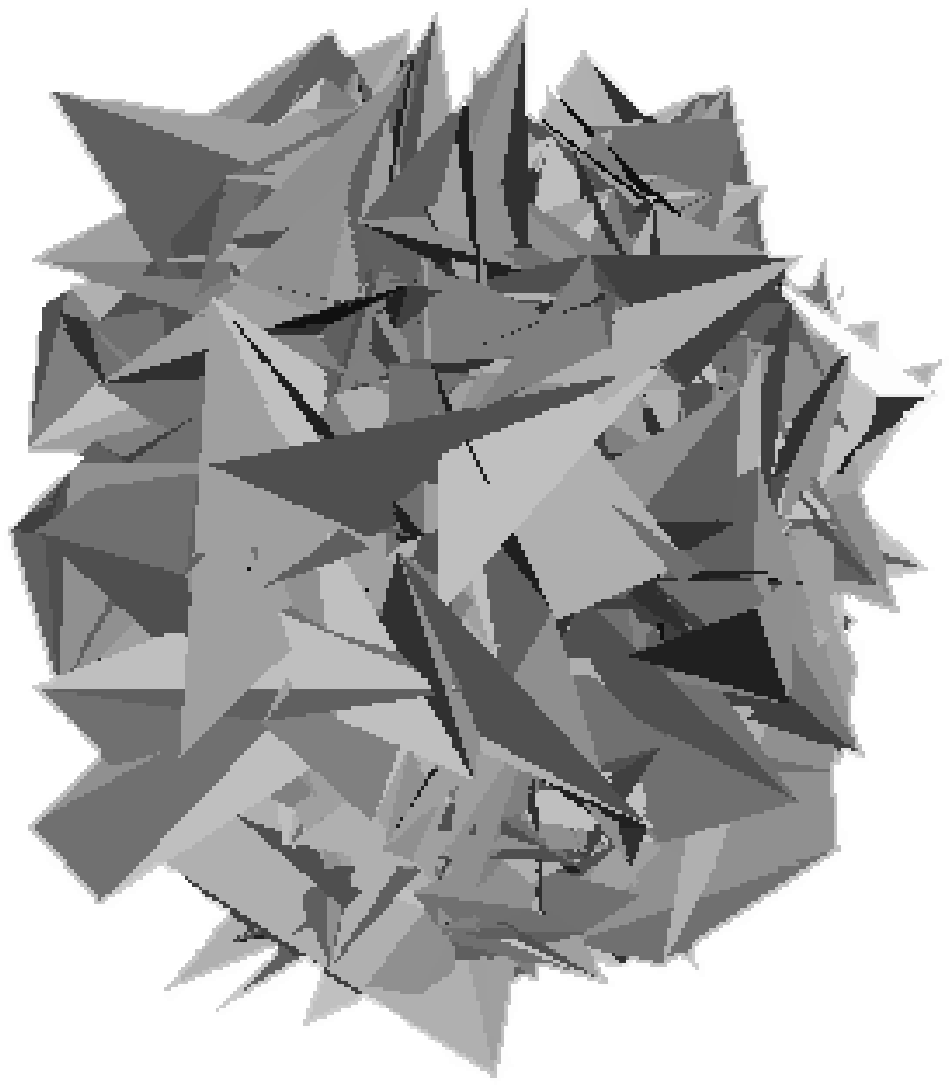,width=6cm,angle=180}}  
\caption{A snapshot of a thermalized configuration ($d=3$) for a {\em phantom}  
membrane ($b=0$), with vanishing bending rigidity, for comparison with   
Fig.\ref{fig__snapshot} ($N=4225$).}  
\label{fig__crumpled}  
\end{figure}  
  
The analysis of the structure function for larger values of $q$ is  
also revealing. The picture emerging so far is that  
a SA fixed-connectivity membrane of linear size $L$ is rough at short distances, of the  
order of $L_{\rm rough}$, but globally flat. The structure function  
must exhibit, in this case, peaks at  
\be\label{flat_modes}  
{\vec q}_{(l_1,l_2)}^{\,L}=\frac{\pi}{L/L_{\rm rough}} (l_1 {\vec G}_1 +  
l_2 {\vec G}_2)  \ , \ l_1,l_2=1, \cdots  
\ee  
where ${\vec G}_{1,2}$ are the standard reciprocal vectors for  
a triangular lattice and the factor $\pi$, as opposed to $2\pi$, is a consequence of free  
rather than periodic boundary conditions. To test Eq.\ref{flat_modes}  
we plot in Fig.\ref{fig__q__peaks} the first (denoted by $n=1$) and  
second (denoted by $n=2$) peaks of the structure function versus $1/L$.   
There is clearly a very good fit.   
Furthermore Eq.\ref{flat_modes} implies that the ratio of wavevectors  
$q^L_{n=2}/q^L_{n=1} = \sqrt{3}$, which is indeed the case: from the fit we  
find   
\be\label{strfnpeaks}  
\frac{b_2}{b_1} = \frac{46}{26} = 1.77 \approx 1.73 = \sqrt{3} \ ,  
\ee  
where $b_n$ is the slope of the fit (see Fig.\ref{fig__q__peaks}).  
From Eq.\ref{flat_modes} we can also extract $L_{\rm rough}$:  
\be\label{struc_rough}  
L_{\rm rough} \approx 8 \ .  
\ee  
This result clearly indicates that one must work with membranes of  
linear size $L$ much larger than $8$ to effectively eliminate  
finite-size effects.  
  
Note that the peaks are damped in intensity   
with increasing n (see Fig.\ref{fig__allstruct}).  
This may be  attributed to the  
membrane having an effective thickness, as given by the roughness exponent.  
  
The previous analysis convincingly establishes that the membrane   
is flat. Further evidence is provided by visualizing typical  
snapshots of membranes after thermalization. This is shown in   
Fig.\ref{fig__snapshot}, where one sees that the membrane is rough at  
short distances ($L_{\rm rough} \approx 8$ by inspection) but  
flat on large scales {--} certainly ``a picture is worth more than  
thousands of numbers.'' In contrast, a snapshot of a thermalized  
configuration, with self-avoidance switched  
off ($b=0$), is shown in Fig.\ref{fig__crumpled}. The dramatic  
effect of self-avoidance is striking. As a final check we performed 
several simulations with a folded initial state and observed the 
subsequent unfolding to the flat phase.    
  
\subsection{BULK DIMENSION $d=4$}\label{Sub_SECT__d4}  
  
\begin{figure}[h]  
\centerline { 
\epsfig{file=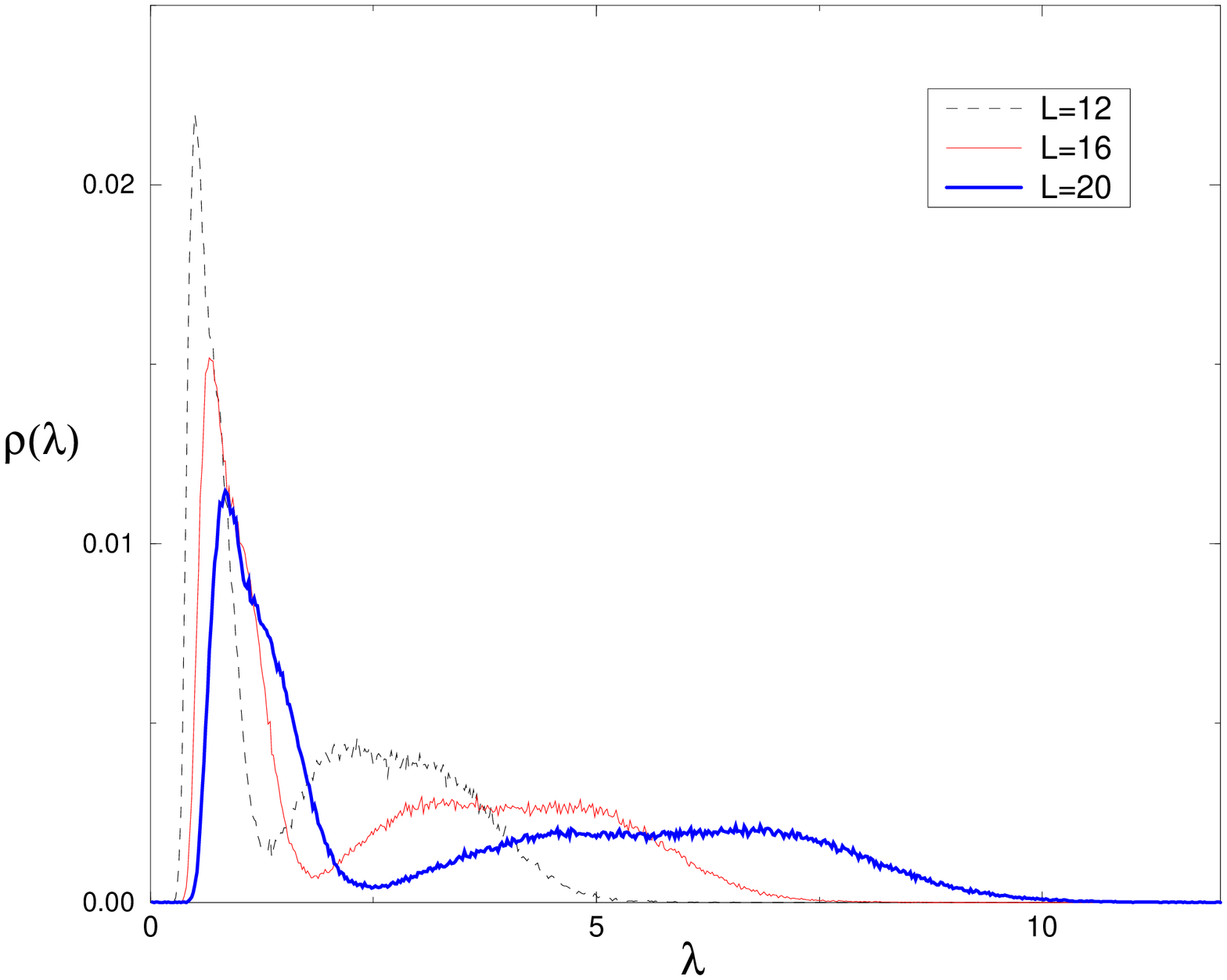,width=4 in,angle=270}}  
\caption{Distribution of eigenvalues for three different volumes   
$N=331$, $N=817$ and $N=1261$ (d=4).}  
\label{fig__no__crumpled}  
\end{figure}

We start by examining the shape eigenvalues Eq.\ref{Inert_ten}.  
In Fig.\ref{fig__no__crumpled} we plot the distribution of eigenvalues for  
different volumes. It is clear from the two-peak structure that a   
crumpled phase can be ruled out. The first peak is   
associated with height fluctuations and the second with size fluctuations.  
These plots are in qualitative agreement with those obtained for  
$d=3$.  
  
\begin{figure}[tb]  
\centerline {\epsfig{file=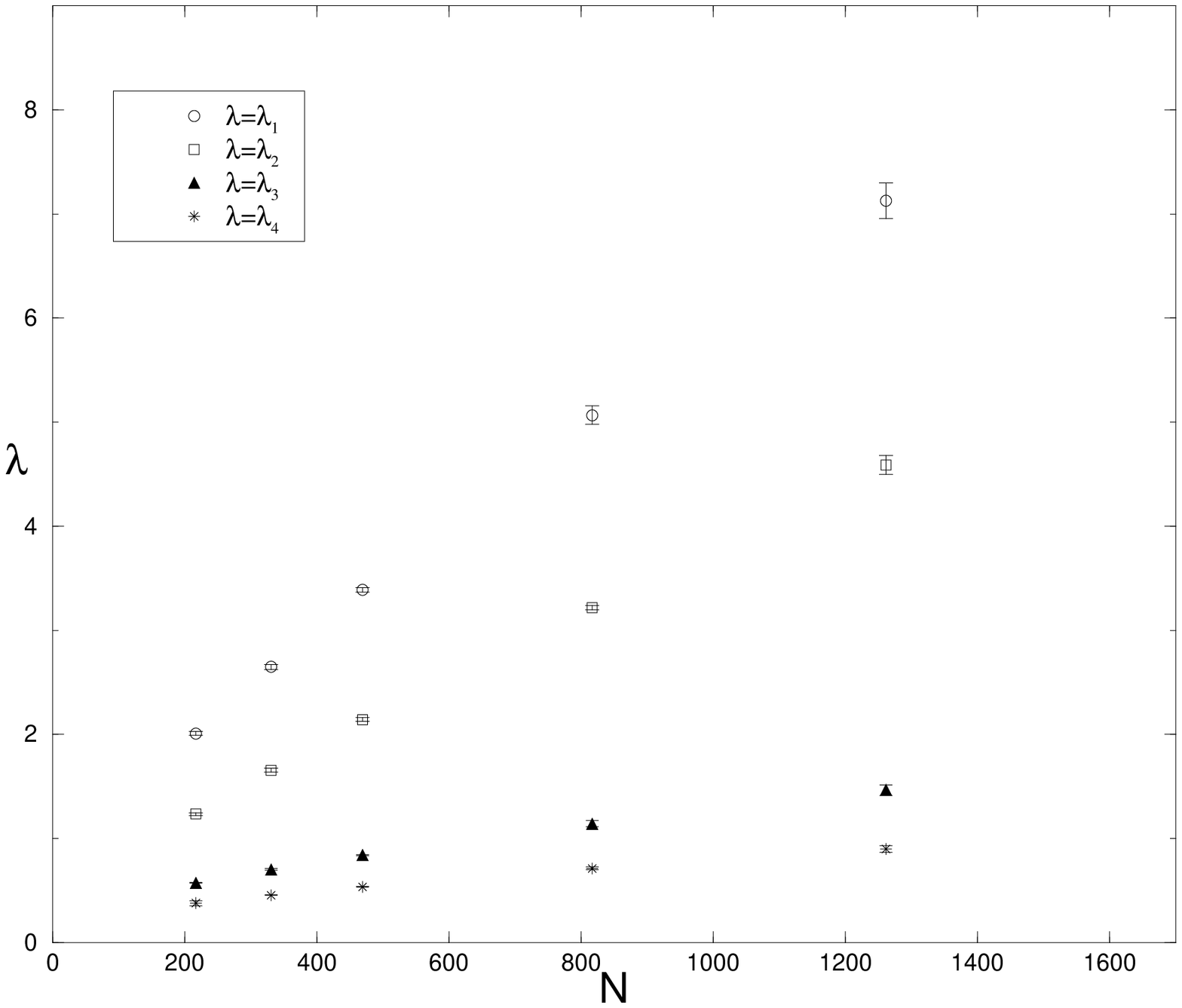,width=10cm,angle=270}}  
\caption{Plot of the eigenvalues as a function of the volume (d=4).} 
\label{fig__vol_size}  
\end{figure}

The expectation values of the eigenvalues are plotted   
as a function of volume in Fig.\ref{fig__vol_size}.    
Both $\lambda_1$ and $\lambda_2$ increase rapidly with volume,   
while $\lambda_3$ and $\lambda_4$ grow slowly.   
  
To compute the exponents $\beta_{\alpha}$, we need to fit the data to the functional  
form in Eq.\ref{Eigen_scl}, which has three free parameters.  
 
Finite size effects are significant for the range of volumes 
we analyzed ($N=127$ to $1261$).   
To evaluate the importance of sub-leading corrections to scaling, 
we once again fit the data to two distinct functional forms. 
The results are shown in Tables \ref{tab__lamb1} and \ref{tab__lamb2}.   
The size exponent spans the range  
\be\label{nu_size4d}  
\nu=0.82(4)-0.90(1) \ ,   
\ee  
 
Although sub-leading corrections significantly affect the extracted 
scaling behavior, it is clear that the size exponent $\nu$ is tending 
to 1. To substantiate this result requires extensive simulations of 
larger volumes. 
   
\begin{table}[htb]  
\centerline{  
\begin{tabular}{|c||c|c|c||c|}  
\hline     
${\bf aN^{\beta}+b}$ & $\beta$ & {$a$}  &  {$b$}   
& {$\chi^2$} \\\hline \hline  
$\lambda_1$       & $0.833(8)$ & $0.0172(9)$  &  $0.47(2)  $   
&  $0.24$ \\\hline  
$\lambda_2$ & $0.81(2)$ & $0.013(2)$  &  $0.20(3) $   
& $0.61$ \\\hline \hline  
$\lambda_3$ & $0.65(1)$ & $0.0119(9)$  &  $0.166(7) $   
& $0.22$ \\\hline  
$\lambda_4$ & $0.620(7)$ & $0.0093(6)$  &  $0.112(5) $   
& $0.10$ \\\hline  
\end{tabular}} 
\caption{Fits to finite-size scaling of the form $aN^{\beta}+b$ for  
the eigenvalues of the shape tensor ($d=4$).}  
\label{tab__lamb1}  
\medskip 
\centerline{  
\begin{tabular}{|c||c|c|c||c|}  
\hline     
${\bf aN^{\beta}+b\log(N)}$& $\beta$ & {$a$}  &  {$b$}   
& {$\chi^2$} \\\hline \hline  
$\lambda_1$       & $0.90(1)$ & $0.010(1)$  &  $0.139(6)  $   
&  $0.48$ \\\hline  
$\lambda_2$ & $0.85(2)$ & $0.0091(2)$  &  $0.060(7) $   
& $0.60$ \\\hline \hline  
$\lambda_3$ & $0.788(6)$ & $0.0038(1)$  &  $0.0574(1) $   
& $0.10$ \\\hline  
$\lambda_4$ & $0.74(2)$ & $0.0031(4)$  &  $0.038(1) $   
& $0.22$ \\\hline  
\end{tabular}}  
\caption{Fits to finite-size scaling of the form $aN^{\beta}+b\log(N)$ for  
the eigenvalues of the shape tensor for (d=4).}  
\label{tab__lamb2}  
\end{table}  
\begin{figure}[ht]  
\centerline {\epsfig{file=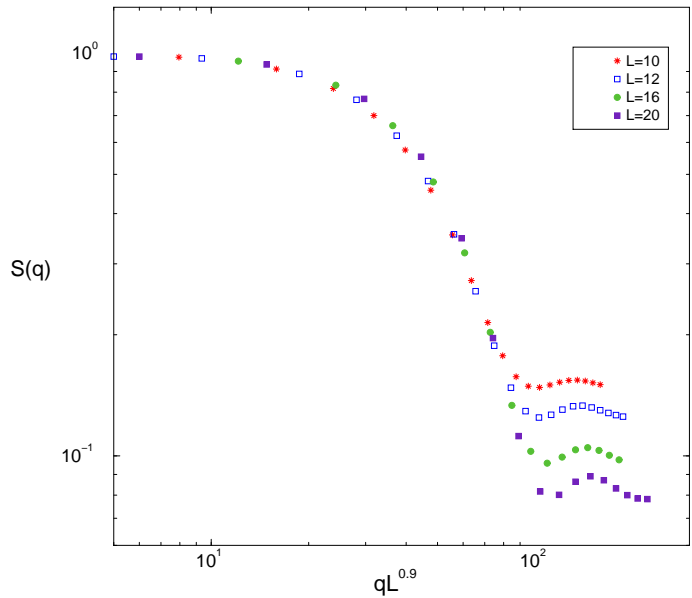,width=8.cm,angle=270}}  
\caption{Log-Log plot of the structure function (d=4) along the direction of 
the maximum eigenvalue of the shape tensor for each volume simulated 
as a function of the scaling variable $qL^{\nu}$ (with $\nu=0.9$).}  
\label{fig__scal_first}  
\end{figure}  
\begin{figure}[bt]  
\centerline {\epsfig{file=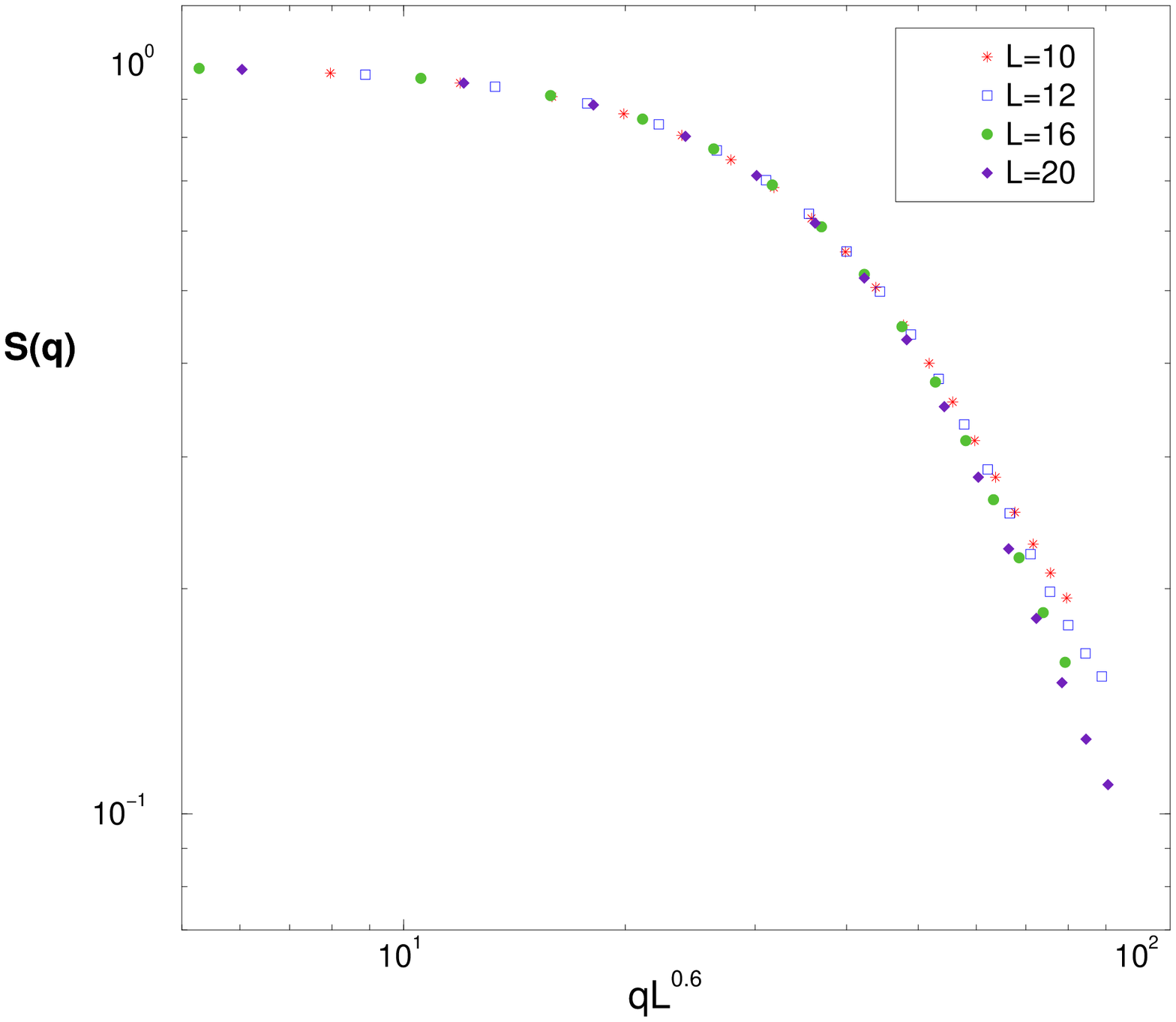,width=8.cm,angle=270}}  
\caption{Log-Log plot of the structure function (d=4) along the direction of 
the minimum eigenvalue of the shape tensor as a function of the 
scaling variable $qL^{\zeta}$ (with $\zeta=0.6$).}   
\label{fig__scal_third}  
\end{figure}

In a flat phase, the scaling of the third and fourth eigenvalue is 
associated with the roughness exponent.  
The roughness exponent falls in the   
range  
\be\label{rough_size4d}  
\zeta=0.65(1)-0.79(1) \ ,  
\ee  
where again, the errors ignore systematic effects. The roughness   
exponent is large, so it implies that the membrane, despite being flat   
(or almost flat) overall, is very rough. For completeness, we quote the   
roughness exponent associated with the smallest eigenvalue; it   
is $\zeta_u=0.62(7)-0.74(2)$. It is somewhat smaller than   
Eq.\ref{rough_size4d} but compatible with it. Whether   
these two exponents are the same is something we cannot establish, and larger  
volume sizes are clearly called for.  
 
The other observable we use to study the properties of the membrane is 
the structure function, defined in Eq.\ref{Struct_fun}. As   
noted earlier, the scaling of the structure function along the 
direction defined   
by the largest eigenvalue is directly related to the size exponent.   
The strategy is to match the structure function as a function of   
the scaled variable $qL^\nu$, where $\nu$ is the size exponent. The   
result of the matching is  
\be\label{nu__match}  
\nu=0.9(1) \ ,  
\ee  
where the error bar needs further explanation. The matching of   
the structure function for small values of $q$ is acceptable for   
values $\nu \geq 0.8$. Near $\nu=0.8$ the range of $q$ for which   
matching is found is somewhat small. This   
range increases with $\nu$, but if $\nu$ is chosen too large (say, 
$\nu >1$), the small-$q$ matching deteriorates and is poor. We note  
that although the matching for $\nu=1$ is still very good, the best 
overall value for $\nu$ is 
$\nu=0.9$.    
The error bar quoted in Eq. \ref{nu__match} is a conservative   
estimate embracing the range of acceptable $\nu$.  
  
We also computed evaluated the structure function along the directions 
of the third and fourth eigenvalue of the shape tensor.   
This yields a roughness exponent (see Fig.\ref{fig__scal_third})    
\be\label{zeta_match}  
\zeta=0.60(7) \ ,  
\ee  
in good agreement with the estimates given in Tables \ref{tab__lamb1} 
and \ref{tab__lamb2}.   
The error bar is a rough estimate indicating the range of exponents 
that yield acceptable scaling.  
  
The situation for the plaquette model for $d > 4$ is much easier to 
analyze. Since a two-dimensional surface almost never self-intersects 
in bulk dimensions five and above, self-avoidance is irrelevant and 
the membrane is always crumpled. The radius of gyration for any $d 
\geq 5$ is given by   
\be\label{R_G_larged}  
R^2_G \sim \log(N) \ ,  
\ee  
which corresponds to $\nu=0$, the Gaussian result.    
 
 
\section{Summary and Conclusions}\label{SECT__conclusions}  
  
\subsection{SUMMARY}  
  
In this section we summarize the most important results obtained from  
the detailed analyses presented earlier.   
The impenetrable plaquette model for $d=3$ is flat for all  
temperatures with critical exponents  
\be\label{final_answer_3d}  
\nu=0.97(4) \ \  \ \zeta=0.63(3) \ ,  
\ee  
with the error reflecting the range of estimates obtained by the  
different methods described. The membrane  
is a flat object at large distances but a very rough one for characteristic  
linear sizes $L \approx 8$. The values for the critical exponents are  
in agreement with the results from BS models  
(see \cite{BowTra:00} for a review). The roughness exponent   
of the flat phase of phantom membranes is given by \cite{BCFTA:96}  
\be\label{phantom_exp}  
\zeta=0.64(2) \ ,  
\ee  
which agrees very well with the result obtained above.   
We conclude that there is a single flat phase for fixed-connectivity  
membranes, describing either a phantom membrane at large bending  
rigidity or a self-avoiding membrane at any non-negative bending  
rigidity.   
  
Embedding dimension $d=4$ has also been studied. Here we find very strong  
evidence that, for all temperatures, there is once again only a flat  
phase, with exponents   
\be\label{final_answer_4d}  
\nu=0.9(1) \ \  \ \zeta=0.65(10) \ .  
\ee  
In this case the quantitative values for the exponent are not as accurate as we  
have for $d=3$ since the lattice volumes considered were not as   
large and finite size effects give rise to systematic errors.   
It would be desirable to simulate larger volumes to narrow   
down the exponent $\nu \sim 1$.    
Gathering all the evidence acquired from the analyses carried out in  
this paper, however, we find it very unlikely that the size exponent  
is not one.  
  
\subsection{CONCLUSIONS}  
  
Our study establishes that the phase diagram of fixed-connectivity membranes  
is very simple and consists only of a flat phase for both $d=3$ and  
$d=4$, and crumpled phases only for $d>4$. In order to rigorously   
compare these theoretical results with experiments, one should   
consider the effect of topological defects \cite{DRNles}, which play  
an important role in any crystalline phase. One would not expect  
defects to alter the long-distance properties of the model, however,    
within the range of temperatures for which the crystalline phase prevails,  
since the overall integrity (unbroken connectivity) of the mesh   
combined with self-avoidance are the key triggers of the flat phase.  
These properties are retained even in the presence of topological  
defects. This overall picture is consistent with the current experimental  
situation {--} no realization of a fixed-connectivity membrane has ever  
been seen in a crumpled state. The observed flat phases have   
exponents consistent with the numerical results of this paper.   
  
Analytical results for SA fixed-connectivity membranes are reviewed in  
\cite{BowTra:00}. Here we simply note that the $\epsilon$-expansion, at  
zero bending rigidity, predicts a unique infra-red stable fixed point  
and no phase transitions, in agreement with our simulations. The  
current value of the size exponent within the $\epsilon$-expansion,  
computed to two-loop order in \cite{DaWi:96,WiDa:97}   
(see \cite{Wiese:00} for a review), however, definitely seems to be  
less than one. It would be of great interest to know if the inclusion  
of higher-orders in the expansion pushes $\nu$ towards the flat value  
of one.    
  
It is also important to remark that the previous results assume repulsive  
potentials among the monomers. When this restriction is dropped and   
attractive forces are considered \cite{AN1:90}, the picture actually   
changes, and compact phases \cite{AK:91} seem to be possible (see also  
\cite{LP:92,GP:94}. This very interesting possibility  
may naturally occur in some systems,   
so it will certainly be the subject of subsequent work.  
  
Our results also have implications for other models. It is well known   
that anisotropic fixed-connectivity membranes possess an intermediate tubular  
phase \cite{RT:95,BFT:97}. Since the tubular phase is, very roughly speaking,   
crumpled in one dimension only, it has been argued that it may survive   
the incorporation of self-avoidance \cite{BG:97,RT:98,BT:99}.   
Anisotropy cannot be readily introduced in our model,   
since if we tune the self-avoidance coupling to zero, we automatically  
sit in the crumpled phase where anisotropy is known to be   
an irrelevant perturbation. Since adding anisotropic extrinsic curvature  
to the self-avoiding membrane will only flatten the membrane even  
more, this cannot produce a tubular phase either.  One  
should therefore consider different discretizations for the self-avoiding case than  
the ones that are suitable for the phantom model discussed in \cite{BFT:97}.  
  
The understanding of the physical properties of fixed-connectivity membranes are  
also very important in constructing realistic models of  
full-fledged cell membranes. The next step in this very exciting  
goal would be to describe a model of a coupled fluid and fixed-connectivity  
membrane. Since one can safely assume (provided no attractive forces are   
present) that the fixed-connectivity membrane is flat, this property   
alone significantly constrains the different effective theories that  
need to be considered. This is just one of the many exciting problems   
that physicists will need to tackle in the near future.

\bigskip  
  
\noindent {\bf Acknowledgements}  
  
\medskip  
  
We acknowledge fruitful discussions with Mehran Kardar.   
The work of MB, AC and AT was supported by the U.S. Department of Energy under  
contract No.~DE-FG02-85ER40237.   
Some computational resources were provided by a grant from NPACI.  
  
\newpage   
  

\end{document}